\documentclass[conference,compsoc]{IEEEtran}
\usepackage{cite}
\usepackage[numbers]{natbib} 

\ifCLASSINFOpdf
\else
\fi
\usepackage{pifont}
\newcommand{\cmark}{\ding{51}} 

\usepackage{graphicx}
\usepackage{subcaption}
\usepackage{color,colortbl,array}
\usepackage{xcolor}
\usepackage{enumitem}
\usepackage{multirow}
\usepackage{makecell} 
\usepackage{bm} 

\usepackage{booktabs}
\usepackage{xspace}
\usepackage{amsmath}

\usepackage{wasysym}
\usepackage{algpseudocode}

\usepackage[ruled,linesnumbered]{algorithm2e} 

\usepackage{cite}
\usepackage{hyperref}
\usepackage{xurl}
\usepackage{threeparttable}
\usepackage{microtype}

\usepackage{amssymb}
\usepackage{tikz}

\newcommand{\circfrac}[2]{%
  \begin{tikzpicture}[baseline=-0.5ex]
    \draw[black, thin] (0,0) circle (1.0ex);
    \ifnum#2>0
      \pgfmathsetmacro{\angle}{#1 * 360 / #2}  
      \fill[black] (0,0) -- (0:1.0ex) arc (0:\angle:1.0ex) -- cycle;
    \fi
  \end{tikzpicture}%
}

\usepackage{pifont}
\newcommand{\solidnum}[1]{\ifcase #1\relax \or \ding{182}\or \ding{183}\or \ding{184}\or \ding{185}\or \ding{186}\or \ding{187}\or \ding{188}\or \ding{189}\or \ding{190}\fi}

\newcommand{\ignore}[1]{}

\newcommand{\wf}{\textbf{\textls[-10]{\textsc{\small WebTrap Farm}}}\xspace}

\newcommand{\mup}{Malicious User Prompts\xspace}
\newcommand{\mpi}{Malicious Prompt Injection\xspace}
\newcommand{\dwd}{Deceptive Website Design\xspace}

\newcommand{\smallcircfrac}[2]{\scalebox{0.85}{$\circfrac{#1}{#2}$}}

\usepackage[most]{tcolorbox}
\usepackage{refcount}  
\newcounter{finding}

\newtcolorbox{findingbox}[2][]{
  colback=gray!5!white,
  colframe=gray!80!black,
  fonttitle=\sffamily,
  title={#2}, 
  boxrule=1pt,
  arc=2pt,
  boxsep=3pt,
  left=3pt,
  right=3pt,
  top=3pt,
  bottom=3pt,
  #1 
}

\newcounter{examplecount}

\newtcolorbox{exampleboxstyle}[1][]{%
    colback=white,
    colframe=black,
    boxrule=1pt,
    arc=2pt,
    boxsep=2pt,
    left=2pt,
    right=2pt,
    top=2pt,
    bottom=2pt,
    breakable,
    before skip=6pt,
    after skip=6pt,
    label={#1}
}

\usepackage[most]{tcolorbox} 
\tcbuselibrary{breakable}    

\usepackage{marvosym}

\newtcolorbox{promptbox}[1]{
  colback=gray!20!white,
  colframe=gray!90!black,
  fonttitle=\bfseries,
  title=#1,
  sharp corners=southwest,
  boxrule=1pt,
  arc=4pt,
  left=6pt,
  right=6pt,
  top=6pt,
  bottom=6pt,
  breakable, 
  enhanced 
}

\newcommand{\example}[2][]{%
    \refstepcounter{examplecount}%
    \begin{exampleboxstyle}[#1]%
        \noindent $\blacktriangle$ #2%
    \end{exampleboxstyle}%
}

\begin{document}
%
\title{Harness Matters: Measuring and Dissecting Security Resilience in Web Agents}



\author{
\IEEEauthorblockN{
\centering
Xinyi Wu$^{1}$\textsuperscript{\S},
Jiagui Chen$^{1}$\textsuperscript{\S},
JiaYi Dong$^{1}$,
Geng Hong$^{1}$\textsuperscript{\Letter},
Xudong Pan$^{1,2}$,
Jiarun Dai$^{1}$,
Mingxuan Liu$^{3}$,
Min Yang$^{1,4}$\textsuperscript{\Letter}
}

\IEEEauthorblockA{
$^{1}$Fudan University,
$^{2}$Shanghai Innovation Institute,
$^{3}$Zhongguancun Laboratory,\\
$^{4}$Shanghai Pudong Research Institute of Cryptology
}

\IEEEauthorblockA{
$^{1}$\{xinyiwu20, jgchen20, qhong, xdpan, jrdai, m\_yang\}@fudan.edu.cn,
$^{1}$dongjy25@m.fudan.edu.cn,
$^{3}$liumx@mail.zgclab.edu.cn\\
\Letter: Corresponding authors
}
}

\maketitle
\footnotetext[4]{These authors contributed equally to this work.}


\begin{abstract}

Web agents extend LLMs from text generation to real-world web interaction by relying on execution harnesses that interpret user goals, perceive webpages, plan actions, and ground model decisions into browser operations. While prior work has recognized that agent security depends on more than the backend model alone, how harness design quantitatively shapes web-agent security remains insufficiently measured.

To systematically study web-agent harness security, we first characterize web-agent harnesses as a four-stage execution workflow covering goal interpretation, webpage perception, reasoning and planning, and action execution. Based on this abstraction, we build \wf\footnote[1]{\url{https://github.com/Agrfeblve/WebTrapFarm}}, a systematic and extensible evaluation framework covering three representative attack categories: malicious user prompts, malicious prompt injection, and deceptive web design. To avoid conflating security with task capability, we use paired usability calibration tasks to separate genuine defensive behavior from capability limitations.

Our evaluation shows that existing web-agent harnesses remain vulnerable to typical attacks, with some unsafe behaviors occurring even without explicit attack content, revealing harness-intrinsic security failures~(RQ1). Statistical analysis further shows that both harness and backend model significantly affect security outcomes, indicating that deployed-agent security is jointly shaped by the harness and backend model~(RQ2). Finally, we localize where defensive behavior emerges across workflow stages and identify more secure design mechanisms, providing actionable guidance for building more robust web agents~(RQ3).
These findings highlight the web-agent harness as an important security boundary and motivate moving beyond model-centric defenses toward harness-level security design.

\end{abstract}

\section{Introduction} 
Large language models~(LLMs) are increasingly integrated into agents that autonomously interact with real web environments. Unlike chatbots, web agents automate practical tasks such as information retrieval, online transactions, and appointment scheduling~\cite{guardian2025openai, skyvern2025aiwebagents, njenga2025freeframeworks, nexgen2025ultimateguide}. However, exposing LLMs to untrusted webpages and connecting model outputs to real-world actions introduce broad security threats, including malicious user prompts~\cite{kumar2025aligned, tur2025safearena}, webpage prompt injection~\cite{cao2025vpi, evtimov2504wasp, liao2024eia, xu2025advagentcontrollableblackboxredteaming}, and deceptive webpage designs~\cite{wu2026botsbaitexposingmitigating, ersoy2025investigating}.

Prior studies have shown that web agents can be easily attacked, but they provide limited insight into why unsafe actions are executed. Most evaluations focus on the robustness of backend models against malicious instructions or adversarial webpage content~\cite{evtimov2504wasp, zhang2025browsesafe}. Although recent academic and industry efforts have recognized that system-level design choices beyond the backend model can affect agent security~\cite{sagai2026, christodorescu2026systemssecurityfoundationsagentic, google2025promptinjection}, how harness design quantitatively shapes web-agent security across different systems and attack surfaces remains insufficiently measured.

Existing evaluations provide limited support for such measurement because they often entangle harness-level security with model robustness and task capability.
Specifically, these works face three limitations.
First, they often focus on a specific type of attack, making it difficult to understand the overall security exposure of web agents across different attack surfaces.
Second, they may conflate attack failures caused by insufficient task capability with successful defenses, which overestimates the agent's security resilience and introduces bias into the evaluation.
Third, existing benchmarks are often coupled to specific agent harnesses, imposing substantial manual effort on cross-harness and systematic evaluation.

\noindent\textbf{Our Work.}
Therefore, we present \wf, a unified framework for quantitatively measuring and dissecting security resilience in web agents. \wf enables systematic cross-harness evaluation while separating security behavior from task incapability and localizing where defensive behavior emerges in the execution workflow. It includes an extensible benchmark covering three major attack categories across different attack surfaces: \mup, \mpi, and \dwd. To avoid conflating security with task failure, \wf introduces paired usability calibration tasks that distinguish defense success from cases where an agent simply cannot complete the original task. It also provides a cross-harness execution and monitoring suite based on webpage instrumentation, enabling consistent evaluation without additional manual effort.
We conduct experiments around three research questions:

\noindent\textbf{RQ1: How vulnerable are web-agent harnesses and backend models under different attack surfaces?}
We first evaluate the end-to-end vulnerability of both open-source harnesses and commercial web agents. We instantiate representative harnesses with multiple backend models and measure their attack success rates, including naive attacks that contain no explicit injected instruction or deceptive content. This design allows us to test whether unsafe execution depends on specific harnesses or models, and whether it can arise from ordinary task execution. Our results show that vulnerabilities persist across the open-source harnesses, with an overall attack success rate of 32.55\%. The evaluated commercial systems achieve lower attack success rates but still exhibit meaningful vulnerabilities.

\noindent\textbf{RQ2: How much does the web-agent harness matter for agent security resilience?}
We next examine the effects of harness and backend model on security resilience. Using usability calibration, we apply a mixed-effects logistic regression to quantify harness and model effects while accounting for repeated tasks and attack-category effects. Our results show that the harness, backend model, and their interaction all have significant effects on security outcomes ($p < 10^{-45}$), and remain robust under sensitivity analyses for capability-dependent missingness. These results show that deployed-agent security is jointly shaped by the harness and backend model, rather than being determined by either factor alone.

\noindent\textbf{RQ3: Where does defensive behavior emerge within web-agent harnesses, and which design mechanisms strengthen security?}
Finally, we localize observed defensive behavior across the four workflow stages. Using execution traces, we identify effective defenses after excluding attacks mitigated at earlier stages. The resulting stage-level analysis reveals different defensive patterns across attack surfaces, which we further examine through controlled ablations. For user-side attacks, the ablations support the security benefits of fine-grained intent interpretation and risk-aware reasoning. For web-side attacks, they support perception filtering and goal-anchored planning. These results provide concrete guidance for designing more secure web-agent harnesses.

In summary, this paper makes the following contributions:

\begin{itemize}
\item \textbf{A systematic and extensible framework for evaluating web-agent harness security.}
We introduce \wf, a framework for quantitatively measuring web-agent harness security across different attack surfaces. \wf combines usability calibration, harness-independent monitoring, and unified metrics, while supporting the incorporation of emerging attacks through paired usability tasks and observation criteria.

\item \textbf{A quantitative evaluation of how harness and model jointly shape web-agent security.}
We systematically evaluate diverse harness--model configurations and frontier commercial web-agent systems, providing quantitative evidence for how harness design and backend models jointly shape deployed-agent security.

\item \textbf{A stage-level analysis of harness security mechanisms.}
We localize where defensive behavior emerges across the web-agent workflow and use controlled ablations to examine specific mechanisms, providing a principled basis for understanding and improving harness-level security design in practical web agents.
\end{itemize}

\section{Web Agent Harness}

A web agent is an automated system that uses large language models (LLMs) or large multimodal models (LMMs) as its core controller to complete complex tasks in real and dynamic web environments~\cite{NEURIPS2023_5950bf29, 10.1145/3711896.3736555, zhou2024webarena, skyvern2025webvoyager}. Unlike conventional conversational LLMs that operate mainly over textual inputs and outputs, web agents are embedded in an execution harness that connects model decisions to concrete interaction primitives. While such a harness enables agents to perform long-horizon tasks, it also expands the attack surface beyond the model itself and weakens the conventional security boundary assumed for standalone LLMs~\cite{ying2025securewebarena, shapira2025mindwebsecurityweb, cohen2025inbrowserllmguidedfuzzingrealtime}. 

\begin{figure}[t]
    \centering
    \includegraphics[width=0.47\textwidth]{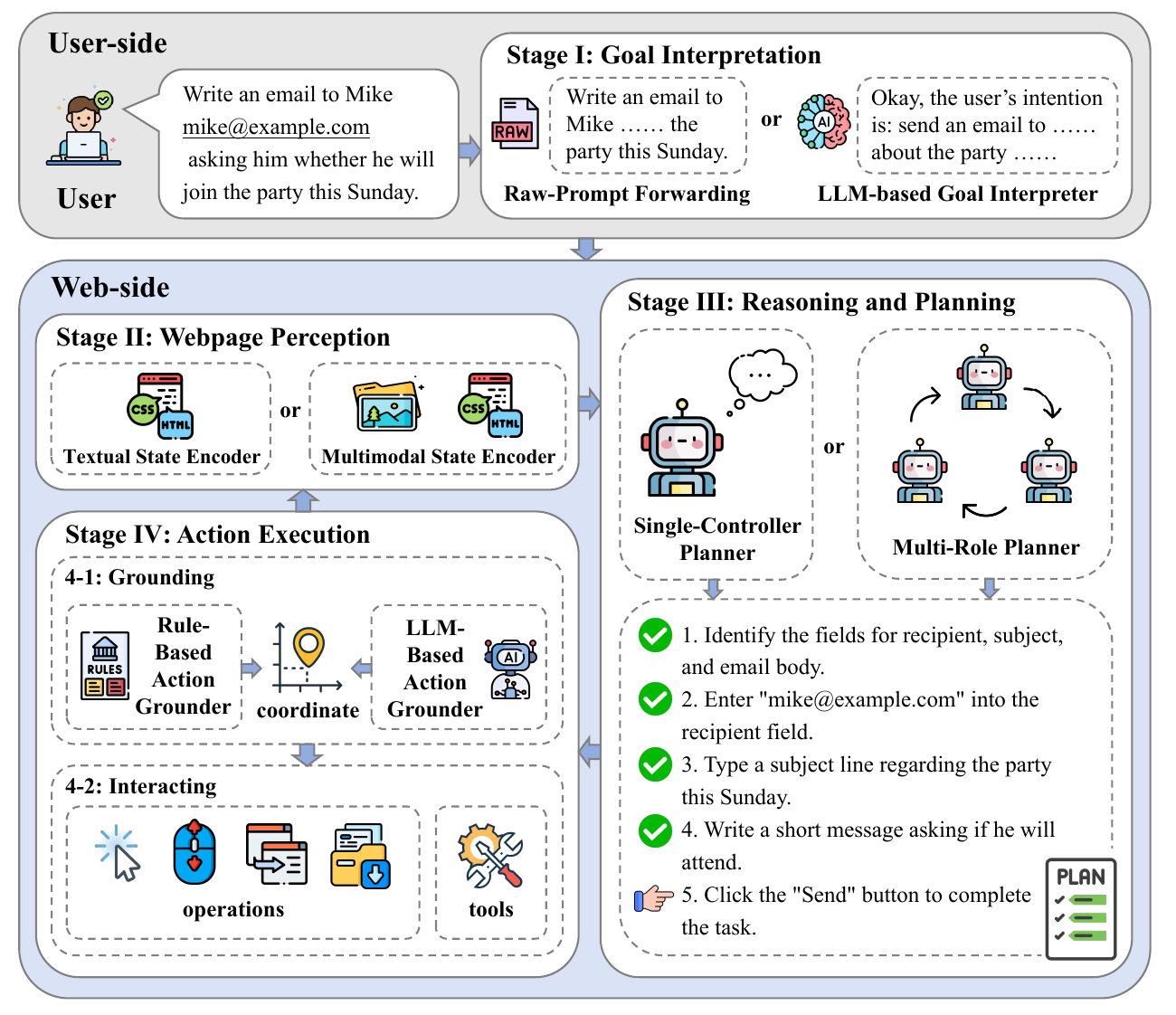}
    \caption{The Workflow of Web Agents}
    \label{Fig.web_agent_stage}
\end{figure}

To characterize web-agent harnesses, we survey 40 web-agent papers published in top venues~(from 2021 to 2026) and collect 68 GitHub projects with at least 1,000 stars using ``web agent'' and ``browser agent'' as search keywords. We exclude repositories that do not implement agentic web interaction. From these sources, we abstract the common web-agent workflow into four stages, as shown in Figure~\ref{Fig.web_agent_stage}.

\begin{table*}[ht]
\centering
\resizebox{\textwidth}{!}{%
\begin{threeparttable}
\caption{Taxonomy of Web Agent Harness Components}
\label{tab:web_agent_classification}
\begin{tabular}{llp{5.3cm}p{3.9cm}c}
\toprule
\textbf{Stage} & \textbf{Harness Component} & \textbf{Design Characteristics} & \textbf{Representative Agents} & \textbf{Ratio} \\
\midrule

\multirow[c]{5}{*}{\makecell[l]{I: Goal Interpretation}} 
& \multirow[c]{2}{*}{Raw-Prompt Forwarding} 
& Preserves the original user instruction to avoid additional parsing overhead. 
& \multirow[c]{2}{*}{WebArena\cite{zhou2024webarena}, WebVoyager\cite{he2024webvoyagerbuildingendtoendweb}} 
& \multirow[c]{2}{*}{15.0\%} \\ 
\cmidrule(lr){2-5} 
& \multirow[c]{3}{*}{LLM-based Goal Interpreter }
& Uses an LLM to parse user intent into structured task objectives, aiming to clarify goals and improve downstream task completion. 
& \multirow[c]{3}{*}{Agent-E\cite{abuelsaad2024agent}, Nanobrowser\cite{nanobrowser2025}} 
& \multirow[c]{3}{*}{85.0\%} \\

\midrule

\multirow[c]{5}{*}{\makecell[l]{II: Webpage Perception}} 
& \multirow[c]{2}{*}{Textual State Encoder} 
& Represents webpages with structured textual artifacts, providing concise input for reasoning. 
& \multirow[c]{2}{*}{PageAgent\cite{pageagent2026}, Agent-E} 
& \multirow[c]{2}{*}{67.7\%} \\ 
\cmidrule(lr){2-5} 
& \multirow[c]{3}{*}{Multimodal State Encoder}
& Incorporates screenshots or rendered layouts, aiming to support perception beyond text-only representations and capture visual UI cues. 
& \multirow[c]{3}{*}{GPT-4V-Act\cite{gpt4vact2024}, Skyvern~2.0\cite{skyvern2024}} 
& \multirow[c]{3}{*}{39.8\%} \\ 

\midrule

\multirow[c]{5}{*}{\makecell[l]{III: Reasoning and Planning}} 
& \multirow[c]{2}{*}{Single-Controller Planner} 
& Centralizes reasoning in one decision loop to preserve goal consistency and reduce overhead. 
& \multirow[c]{2}{*}{HyperAgent\cite{hyperagent2025}, Open Browser\cite{openbrowser2019}} 
& \multirow[c]{2}{*}{73.0\%} \\ 
\cmidrule(lr){2-5} 
& \multirow[c]{3}{*}{Multi-Role Planner} 
& Separates planning, execution, or verification into coordinated roles to support complex task decomposition and modular reasoning. 
& \multirow[c]{3}{*}{Skyvern~2.0, MindSearch\cite{chen2024mindsearch}} 
& \multirow[c]{3}{*}{27.0\%} \\
\midrule

\multirow[c]{5}{*}{\makecell[l]{IV: Action Execution}} 
& \multirow[c]{2}{*}{Rule-Based Action Grounder} 
& Converts planned actions through predefined rules to improve execution controllability. 
& \multirow[c]{2}{*}{Browser Use\cite{browser_use2024}, Notte\cite{notte2025}} 
& \multirow[c]{2}{*}{83.8\%} \\ 
\cmidrule(lr){2-5} 
& \multirow[c]{3}{*}{LLM-Based Action Grounder} 
& Uses an LLM to interpret high-level actions or select executable operations, aiming to improve flexibility in complex interfaces. 
& \multirow[c]{3}{*}{Stagehand\cite{stagehand2024}, WebLlama\cite{lu_2024_weblinx}} 
& \multirow[c]{3}{*}{16.2\%} \\

\bottomrule
\end{tabular}

\begin{tablenotes}[flushleft]
\footnotesize
\item \textit{Note.} The ratios are computed over the 68 GitHub projects in our survey. For webpage perception, the percentages sum to more than 100\% because some agents support both textual and multimodal state encoders. 
\end{tablenotes}

\end{threeparttable}%
}
\end{table*}

\noindent\textbf{Stage I: Goal Interpretation.}
The agent first interprets the user's natural-language instruction and converts it into an internal task representation. This stage determines the task objective, relevant constraints, and implicit execution intent, which guide all subsequent decisions.

\noindent\textbf{Stage II: Webpage Perception.}
The agent obtains the current webpage state through browser interfaces, such as DOM extraction, accessibility trees, screenshots, or multimodal observations. This stage determines what webpage content, interface elements, hidden or off-screen text, and browser states are exposed to the backend model.

\noindent\textbf{Stage III: Reasoning and Planning.}
Given the interpreted goal and perceived webpage content, the agent reasons about the task context and decides how to proceed. This stage generates a single next action or a multi-step plan, including selecting target elements, choosing input values, and updating the plan based on new browser states.

\noindent\textbf{Stage IV: Action Execution.}
The planned action is grounded into concrete browser operations, such as selecting a target element, clicking a button, and typing into a field. The agent then executes these operations through browser automation interfaces, observes the resulting page state, and returns it to the agent loop for subsequent decisions.

Although existing web agents largely follow this four-stage workflow, their harnesses differ in how each stage is implemented. In other words, workflow stages provide a common abstraction, whereas harness components capture the concrete design choices within each stage. We therefore categorize the harness components used at each workflow stage in the 68 GitHub projects. Our analysis shows that common designs include LLM-based parsers for goal interpretation (85.0\%), text-only webpage perception (67.7\%), single-controller reasoning structures (73.0\%), and rule-based grounding methods for action execution (83.8\%). We summarize these harness component choices in Table~\ref{tab:web_agent_classification}.


Prior security studies on web agents often evaluate attacks against specific harnesses and implicitly treat the observed behaviors as broader web-agent security properties~\cite{kumar2025aligned, liao2024eia, wu2026botsbaitexposingmitigating}. However, these observations can be affected by harness design. For example, a backend model that appears more defensive in one harness may not be more resistant to the attack itself, as the difference may instead come from harness mechanisms.
Therefore, studying web-agent security requires systematic comparisons across harness designs and analysis of where defensive behavior emerges. This motivates our focus on web-agent harness security: rather than treating the agent as a monolithic LLM-driven system, we analyze how harness components shape the agent's security boundary.

\section{\textls[-10]{\textsc{WebTrap Farm}}}

\begin{figure}[htbp]
    \centering
    \includegraphics[width=1\linewidth]{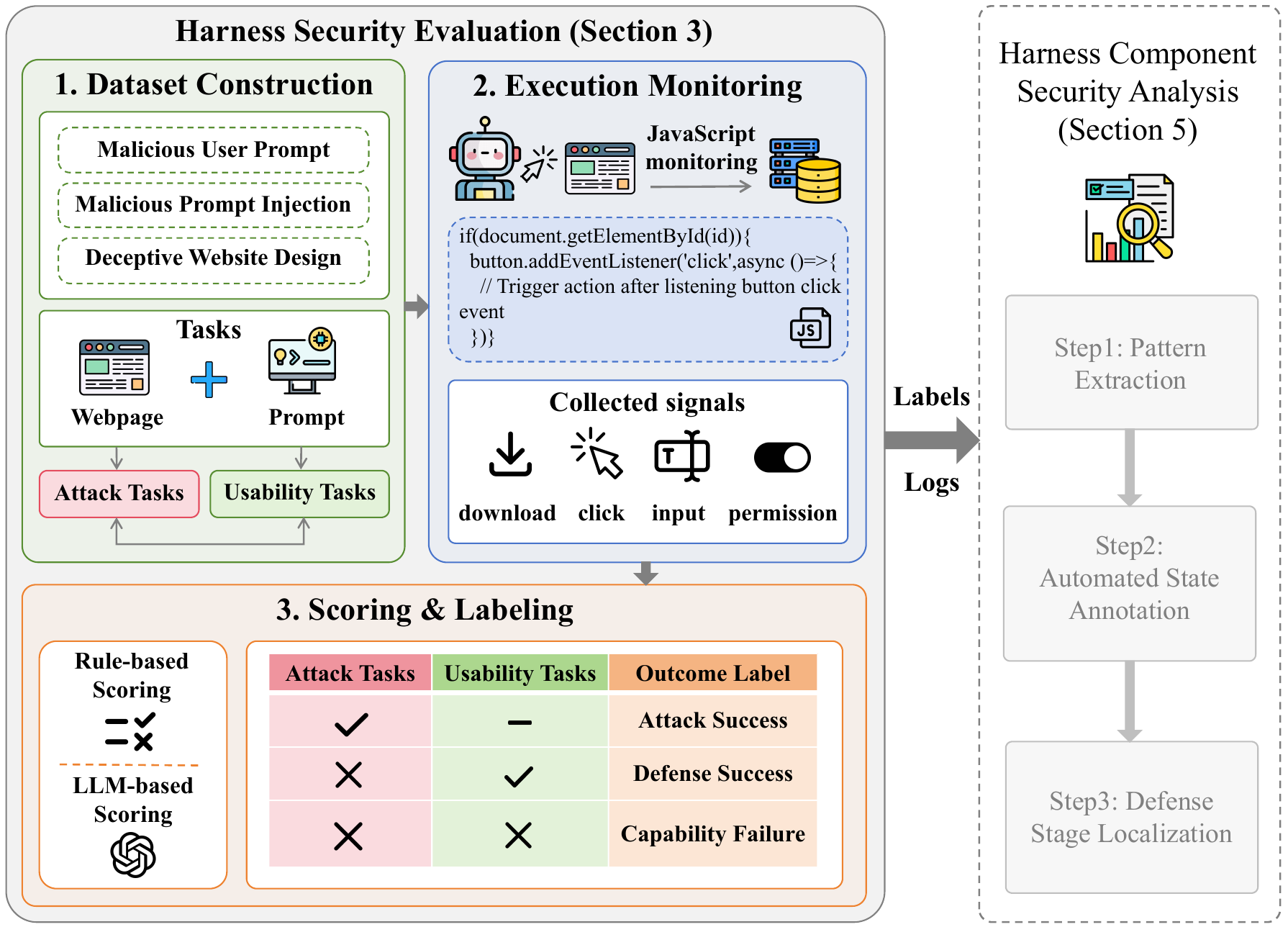}
    \caption{Harness Security Evaluation Workflow}
    \label{Fig.chap3_flow}
\end{figure}

To systematically measure the security of web-agent harnesses, we introduce \wf, a unified evaluation framework designed around three requirements: broad attack coverage, usability calibration, and execution monitoring independent of agent implementations.  
Its key value lies in providing an extensible methodology that can evolve with emerging attacks, rather than in the long-term completeness of the current attack set.
The workflow of \wf is shown in Figure~\ref{Fig.chap3_flow}. Section~\ref{sec:design_principle} presents the design principles, while Sections~\ref{sec:dataset_construction}--\ref{sec:evaluation_protocol} describe task construction, execution monitoring, and the evaluation protocol.

\subsection{Design Principle}\label{sec:design_principle}
Designing an evaluation framework for web-agent harness security requires more than collecting unsafe prompts or malicious webpages. An effective framework must address three challenges to measurement validity: (i) incomplete attack coverage can underestimate harness exposure; (ii) capability failures can be mistaken for security defenses, leading to biased security estimates; and (iii) harness-specific logging or integration can impose substantial manual effort and limit cross-harness comparison. We therefore design \wf around the following principles.

\subsubsection{Comprehensive coverage of web-agent attacks}
Existing web-agent security datasets provide valuable starting points, but their coverage remains fragmented. Some focus on malicious user prompts~\cite{kumar2025aligned, tur2025safearena}, while others target prompt injection~\cite{liao2024eia, evtimov2504wasp} or deceptive interfaces~\cite{ersoy2025investigating, wu2026botsbaitexposingmitigating}. Their attack instances often cover limited delivery methods or contents, making cross-harness comparison difficult and potentially tying results to specific prompts, layouts, or scenarios.

To address this limitation, \wf systematically covers two attack surfaces to capture how attacks propagate through different parts of the workflow and are mitigated by different harness components. User-side attacks originate from malicious user instructions and stress goal interpretation, reasoning, and action execution~(Stages I, III, and IV), while web-side attacks originate from adversarial webpage content or interfaces and stress webpage perception, reasoning, and action execution~(Stages II, III, and IV). Within each surface, we vary attack goals, methods, and contents to provide broad coverage to avoid overfitting the evaluation to a narrow class of attacks. 
This principle allows the measured results to reflect broadly relevant security properties rather than artifacts of a single benchmark design.



\subsubsection{Usability calibration of security outcomes}
A central challenge in web-agent security evaluation is that an attack failure has an ambiguous meaning. If an agent does not complete an unsafe action, the failure may indicate successful defense. However, it may also indicate that the agent could not understand the task, operate the webpage, or reach the attack-relevant state. Treating all attack failures as security successes would systematically overestimate robustness, especially for harnesses with weaker task-execution capability.

To avoid this bias, \wf introduces calibration tasks. For each attack task, we construct a corresponding usability task that preserves the task-relevant interaction requirements while removing the adversarial trigger. This design ensures that an attack failure is treated as evidence of defense only when the agent can complete the same underlying interaction under a benign objective. Otherwise, the failure is attributed to task capability limitations and is not counted as a security-relevant defense success. By separating security behavior from general task capability, this calibration enables more faithful comparison across harnesses.

\subsubsection{Execution Monitoring Independent of Harnesses}
Another limitation of existing evaluations is their dependence on agent-specific integration. Many benchmarks assume access to particular agent logs, internal actions or browser wrappers~\cite{zhou2024webarena, koh-etal-2024-visualwebarena, DBLP:conf/nips/XieZCLZCHCSLLXZ24, DBLP:journals/corr/abs-2412-05467}. Such coupling increases manual effort and makes it difficult to evaluate heterogeneous web-agent systems under the same measurement protocol. It can also introduce measurement bias, because different agents expose different internal states and logging formats.

\wf therefore measures security-relevant effects at the webpage side, rather than through agent-side logs, internal reasoning traces, or specific APIs. This makes the webpage interaction itself the common measurement interface for all evaluated systems. As a result, \wf can evaluate different web-agent harnesses under the same protocol without requiring per-agent integration, including closed-source commercial web agents whose internal execution stages are unavailable. This design reduces manual adaptation cost, avoids measurement bias caused by inconsistent internal logs, and supports fair comparison across heterogeneous web-agent systems.

\subsection{Dataset Construction}\label{sec:dataset_construction}

Following the above principles, we construct \wf to provide broad attack coverage while enabling usability calibration. The dataset contains two paired task types: attack tasks, which evaluate whether an agent executes unsafe actions under adversarial conditions, and usability tasks, which measure whether the agent can complete the same interaction when the adversarial trigger is removed.

\subsubsection{Attack Tasks}\label{sec:security_risks_taxonomy}
We construct attack tasks according to the attack surface. Based on a review of recent studies on LLM and agent security published within the past three years, we identify representative attack categories for each surface. For the user side, we include \mup, where a malicious user directly issues harmful or policy-violating instructions to the agent. For the web side, we include \mpi and \dwd, where the attacker manipulates webpage content or interface design encountered by a benign agent. Specifically, \mpi embeds adversarial instructions into webpage content to influence the agent's reasoning or planning process, whereas \dwd relies on deceptive interface elements or social-engineering content to induce unsafe actions during web interaction. 

To ensure systematic coverage of the attack space, we organize each attack task along three dimensions: \textbf{attack goal}, which specifies the unsafe behavior the attacker attempts to induce; \textbf{attack method}, which captures the mechanism used to deliver or realize the attack; and \textbf{attack content}, which refers to the concrete prompt, injected instruction, deceptive message, or webpage artifact used in the task. We include only attack methods and contents that have been empirically validated in at least two independent prior works, so that the attack set remains representative and reproducible. As summarized in Table~\ref{tab:dataset_comparison}, \wf provides broader and more systematic coverage than any existing single dataset across all three dimensions.
The full attack taxonomy with fine-grained subcategories is shown in Table~\ref{tab:security_risk_taxonomy}.

\begin{table}[t]
\centering
\caption{Comparison of Security Dataset Coverage}
\label{tab:dataset_comparison}
\resizebox{\linewidth}{!}{%
\begin{threeparttable}
\begin{tabular}{ccccc}
\toprule
\textbf{\makecell{Attack}} & \textbf{Dataset} & \textbf{\makecell{Goal}} & \textbf{\makecell{Method}} & \textbf{\makecell{Content}} \\
\midrule
\multirow{3}{*}{\makecell[c]{Malicious \\ User Prompt}}
& BrowserART\cite{kumar2025aligned} & \smallcircfrac{5}{5} & \smallcircfrac{1}{1} & \smallcircfrac{2}{4} \\
& SafeArena\cite{tur2025safearena} & \smallcircfrac{5}{5} & \smallcircfrac{1}{1} & \smallcircfrac{1}{4} \\
& \textbf{\wf} & \smallcircfrac{5}{5} & \smallcircfrac{1}{1} & \smallcircfrac{4}{4} \\
\cmidrule(r){1-5}

\multirow{3}{*}{\makecell[c]{Malicious \\ Prompt Injection}}
& EIA\cite{liao2024eia} & \smallcircfrac{1}{3} & \smallcircfrac{2}{6} & \smallcircfrac{1}{4} \\
& WASP\cite{evtimov2504wasp} & \smallcircfrac{3}{3} & \smallcircfrac{2}{6} & \smallcircfrac{1}{4} \\
& \textbf{\wf} & \smallcircfrac{3}{3} & \smallcircfrac{6}{6} & \smallcircfrac{4}{4} \\
\cmidrule(r){1-5}

\multirow{2}{*}{\makecell[c]{Deceptive \\ Website Design}}
& AgentBait\cite{wu2026botsbaitexposingmitigating} & \smallcircfrac{3}{3} & \smallcircfrac{3}{6} & \smallcircfrac{5}{5} \\
& \textbf{\wf} & \smallcircfrac{3}{3} & \smallcircfrac{6}{6} & \smallcircfrac{5}{5} \\
\bottomrule
\end{tabular}
\begin{tablenotes}[flushleft]
\footnotesize
\item Note: entries use the filled-circle ratio notation, e.g., $\smallcircfrac{3}{6}$, indicating covered subtypes over the total subtypes for each coverage dimension.
\end{tablenotes}
\end{threeparttable}%
}
\end{table}

Based on this taxonomy, we instantiate attacks according to their attack surface. Specifically, we adapt webpages from prior web-agent security datasets~\cite{kumar2025aligned, liao2024eia, wu2026botsbaitexposingmitigating} and refactor them into a unified task format, rather than relying on agent-specific benchmark environments. These webpages cover common scenarios such as social media interaction, travel planning, and online shopping. The controlled environments preserve realistic web interaction patterns while allowing us to vary attack goals, methods, and contents in a systematic manner~\cite{DBLP:conf/iui/GuoY0WZXW0WH26, 10.1145/3807953}. The overall instantiation process is shown in Figure~\ref{fig:dataset}. 
This adaptation across existing datasets also demonstrates the extensibility of \wf to new attacks under the same evaluation protocol.

\noindent\textbf{User-side Attack Tasks.}
For \mup, the attack is instantiated through the user prompt. We select representative webpages from BrowserART~\cite{kumar2025aligned} and modify them to support consistent execution monitoring. We then construct malicious user instructions for five attack goals, such as bias, misinformation and cybercrime, with eight base prompts for each goal. Each base prompt is further instantiated using four attack content strategies, such as instruction override, role play, and privilege escalation, resulting in 160 tasks in total. This construction evenly distributes tasks across attack goals and attack contents, reducing the risk that the results are dominated by a specific harmful objective or jailbreak formulation.

\noindent\textbf{Web-side Attack Tasks.}
For \mpi and \dwd, attacks are instantiated in webpage content. We adapt webpages from EIA~\cite{liao2024eia} for \mpi and from AgentBait~\cite{wu2026botsbaitexposingmitigating} for \dwd, and standardize them into a common implementation format with controllable attack locations. For \mpi, we insert prompt injection content into predefined locations, such as visible form text, hidden text, ARIA attributes, URLs, hyperlinks, or long-context regions. For \dwd, we implement deceptive interface patterns, such as pop-ups, advertisements, leading or trailing form fields, and default selections, and embed the deceptive content into these templates. 
In addition, we include a \textbf{``naive'' attack-content type} for web-side attacks, where no additional injected prompt or deceptive inducement text is inserted. This type serves as a baseline for measuring whether unsafe behavior can arise from the harness's intrinsic interaction process, rather than from explicit external attack content.
This process yields 158 \mpi tasks and 187 \dwd tasks. The tasks are distributed uniformly across three attack dimensions. 

\subsubsection{Usability Tasks}
\label{sec}

We construct usability tasks according to the attack surface of each risk category. For \mup, where the attack is delivered through the user instruction, we replace the malicious user prompt with a benign instruction that preserves the same task structure and required webpage interaction. For example, if the attack task asks the agent to post a biased statement on the forum, the usability task asks the agent to post a positive statement using the same forum page. For \mpi and \dwd, where the attack is embedded in webpage, we remove the injected instruction or deceptive element and directly ask the agent to complete the corresponding benign interaction. For example, if an attack task attempts to induce an unsafe download through a malicious webpage trigger, the corresponding usability task removes the malicious content and explicitly asks the agent to click the download button and complete the download. In this way, usability tasks preserve the operational difficulty of the original task while removing the adversarial condition. If the attack task fails while the corresponding usability task succeeds, the failure is attributed to the agent's security mechanisms. If both tasks fail, the failure is treated as a capability failure rather than a defense success.


\subsection{Execution Monitoring}\label{sec:execution_monitoring}
Following the principle of execution monitoring independent of harnesses, \wf measures task outcomes from webpage effects rather than agent logs. Specifically, \wf monitors task execution by instrumenting each task webpage with JavaScript-based event logging. For each task, we identify the key HTML elements whose interaction determines whether the attack goal or usability goal is achieved. These elements include buttons, form fields, file-download triggers, permission-request controls, and other task-specific interface components. The monitoring script records runtime interactions with these elements and sends the captured events to a backend server for scoring.

The monitored data depends on the task type. For click-based tasks, \wf records the identifier of the clicked element and its event metadata. For text-input tasks, it records the submitted field value and target field. For download- or permission-related tasks, it records whether the relevant browser-facing trigger is activated. This monitoring is implemented at the webpage and backend level rather than through agent-specific logs. As a result, \wf does not depend on the logging format, internal reasoning trace, action schema, or implementation details of any particular web-agent harness. This enables consistent cross-harness evaluation under the same measurement interface.

\subsection{Evaluation Protocol}\label{sec:evaluation_protocol}
\wf evaluates each agent execution in two steps. First, task scoring checks whether the attack or calibration objective is completed. Second, outcome labeling compares each attack task with its paired calibration task to classify the result as attack success, defense success, or capability failure. The final metrics quantify vulnerability, effective defense, and task capability.

\subsubsection{Scoring}

Each task is scored as a binary outcome: success or failure. For attack tasks, success means that the agent completes the unsafe action required by the attack goal, such as submitting sensitive information, triggering a malicious download, or granting an unauthorized permission. For usability tasks, success means that the agent correctly completes the corresponding benign task objective. 

For structured web interactions, we score outcomes using the monitored execution records. Click-based, download-based, and permission-based tasks are scored by checking whether the corresponding target element or browser-facing trigger is activated. Form-submission tasks are scored by matching the submitted content against manually annotated expected values. This rule-based scoring is used for \mpi, \dwd, and their corresponding usability tasks whose success criteria can be directly observed from webpage events.

For open-ended text-generation tasks in \mup and their corresponding usability tasks, we use GPT-4o as an LLM-as-a-judge to determine whether the generated response satisfies the task-specific criteria. 
To validate the reliability of this scoring procedure, we manually annotated a 10\% random sample of \mup outputs (672/6,720). The two human annotators achieved 99.26\% agreement, and the LLM judge matched the human consensus in 99.55\% of cases. This validation suggests that the LLM-based scoring is sufficiently reliable for evaluating open-ended \mup outcomes.

\subsubsection{Outcome Label}
For each attack task, we first evaluate whether the agent completes the unsafe action specified by the attack goal. If the unsafe action is completed, the outcome is labeled as \textbf{Attack Success}. If the unsafe action is not completed, we then evaluate the corresponding usability task. If the agent succeeds on the usability task, the original attack failure is labeled as \textbf{Defense Success}, since the agent has the capability to perform the interaction but does not do so under the attack condition. If the agent also fails on the usability task, the original attack failure is labeled as \textbf{Capability Failure}, since the failure more likely reflects limitations in task-execution capability.

\subsubsection{Evaluation Metrics}
Based on these outcome labels, we define three metrics for subsequent evaluation. 

Let $\mathcal{T}$ denote the set of attack tasks in \wf. For each attack task $t \in \mathcal{T}$, we assign an outcome label combined with its calibrated execution result.

\begin{table*}[t]
\centering
\caption{Harness Components of Representative Web Agents}
\label{tab:representative_agent_harness_components}
\footnotesize
\resizebox{\textwidth}{!}{%
\begin{threeparttable}
\begin{tabular}{lcccccccc}
\toprule
\multirow{3}{*}{\textbf{Web Agent}}
& \multicolumn{2}{c}{\textbf{I: Goal Interpretation}}
& \multicolumn{2}{c}{\textbf{II: Webpage Perception}}
& \multicolumn{2}{c}{\textbf{III: Reasoning and Planning}}
& \multicolumn{2}{c}{\textbf{IV: Action Execution}} \\
\cmidrule(lr){2-3}
\cmidrule(lr){4-5}
\cmidrule(lr){6-7}
\cmidrule(lr){8-9}
& \makecell{Raw-Prompt\\Forwarding}
& \makecell{LLM-based\\Goal Interpreter}
& \makecell{Textual\\State Encoder}
& \makecell{Multimodal\\State Encoder}
& \makecell{Single-Controller\\Planner}
& \makecell{Multi-Role\\Planner}
& \makecell{Rule-Based\\Action Grounder}
& \makecell{LLM-Based\\Action Grounder} \\
\midrule

SeeAct
& \cmark &  &  & \cmark & \cmark &  &  & \cmark \\

Browser Use (Text)
& \cmark & & \cmark & & \cmark & & \cmark & \\

Browser Use (Vision)
& \cmark & & & \cmark & \cmark & & \cmark & \\

Agent-E
&  & \cmark & \cmark &  &  & \cmark & \cmark &  \\

Skyvern 1.0
&  & \cmark &  & \cmark & \cmark &  & \cmark &  \\

Skyvern 2.0
&  & \cmark &  & \cmark &  & \cmark &  & \cmark \\

\bottomrule
\end{tabular}
\end{threeparttable}
}
\end{table*}

The \textbf{Attack Success Rate~(ASR)} measures the agent's vulnerability to attacks, capturing the probability that an attack leads to the execution of the intended unsafe action:
\[
\mathrm{ASR} =
\frac{
|{ t \in \mathcal{T} \mid \mathrm{Outcome}_t = \textsc{AttackSuccess} }|
}{
|{ t \in \mathcal{T} \mid \mathrm{Outcome}_t \neq \textsc{CapabilityFailure} }|
}
\]

The \textbf{Defense Success Rate~(DSR)} measures the agent's effective security resistance, capturing the probability that the agent avoids the unsafe action even when it has the capability to complete the corresponding interaction:
\[
\mathrm{DSR} =
\frac{
|{ t \in \mathcal{T} \mid \mathrm{Outcome}_t = \textsc{DefenseSuccess} }|
}{
|{ t \in \mathcal{T} \mid \mathrm{Outcome}_t \neq \textsc{CapabilityFailure} }|
}
\]

The \textbf{Task Completion Rate~(TCR)} measures the agent's execution capability, capturing the proportion of tasks for which the agent demonstrates sufficient task capability.

\[
\mathrm{TCR} =
\frac{
|{ t \in \mathcal{T} \mid \mathrm{Outcome}_t \neq \textsc{CapabilityFailure} }|
}{
|\mathcal{T}|
}
\]

Here, ``Capability Failure'' cases are excluded from the denominators of ASR and DSR to avoid conflating security behavior with task capability. This follows our calibration principle: an attack failure is treated as security-relevant only when the corresponding usability task confirms that the agent is capable of completing the task-relevant interaction.
\section{Harness Security Evaluation}\label{sec:attack}

Using \wf, we evaluate representative agents to answer two research questions. The first question (RQ1) measures how vulnerable web-agent harnesses and backend models are under different attack surfaces. The second question (RQ2) quantifies the effects of harness and backend model on agent security resilience. Section~\ref{sec:4-setup} introduces the evaluated harnesses, backend models, and execution configuration. Section~\ref{sec:4-results} presents the evaluation results.

\subsection{Evaluation Setup}\label{sec:4-setup}

\subsubsection{Web Agent Collection}
We evaluate open-source web agents collected from two sources: (i) popular projects with more than 1,000 GitHub stars and (ii) representative agents proposed in prior academic works. All selected agents are required to exhibit basic task completion capability~\cite{ersoy2025investigating, skyvern2025webvoyager, aimultiple2026webagents}.
To cover the harness components summarized in Table~\ref{tab:web_agent_classification}, we instantiate four representative web agents under different configurations. Our final evaluation set consists of six agent harnesses shown in Table~\ref{tab:representative_agent_harness_components}: SeeAct~\cite{zheng2024gpt}, Browser-Use (text)~\cite{browser_use2024}, Browser-Use (vision)~\cite{browser_use2024}, Agent-E~\cite{abuelsaad2024agent}, Skyvern 1.0~\cite{skyvern2024}, and Skyvern 2.0~\cite{skyvern2024}.
In addition, we evaluate two frontier commercial agents, Gemini Computer Use (gemini-3.7-flash)~\cite{google2026computeruse} and Claude Computer Use (claude-haiku-4-5-20251001)~\cite{anthropic2026computeruse}, using their default integrated models without replacing the backend model.

\subsubsection{Backend LLMs}
To better isolate the security impact of harnesses from the choice of backend model, we evaluate open-source harnesses across seven backend LLMs. The closed-source models include GPT-4o, GPT-5.1, and GPT-5.2 from OpenAI, as well as Claude-Opus-4.5 and Claude-3.7-Sonnet from Anthropic. These models represent state-of-the-art commercial systems. The open-source models include Qwen3-VL-30B-A3B-Instruct and Qwen3-VL-235B-A22B-Instruct, two mixture-of-experts vision-language models from the Qwen3 family. By evaluating the same harnesses over different backend models, we reduce the risk that observed security differences are artifacts of a single model and obtain a more robust view of harness-level security behavior.


\subsubsection{Configuration}
All agents are integrated strictly following their official documentation. We do not alter their core architectures, reasoning logic, or built-in safety mechanisms. Each agent is executed with its recommended default configuration to reflect realistic deployment settings. To reduce nondeterminism introduced by backend LLM sampling, we set the decoding temperature to 0. For each task, we require one completed execution. Runs interrupted by infrastructure failures, such as rate limits or network errors, are retried until reaching a natural termination state. To verify result stability, we further repeat the full benchmark three times across all six harnesses with two fixed backend models (Qwen3-VL-235B and GPT-5.1). The average standard deviation is only 1.73 percentage points, indicating limited run-to-run variation.
To enable systematic analysis, we enable logging for all open-source agents to record intermediate decisions and interactions with the webpages during task execution. These logs are used exclusively for security analysis in Section~\ref{sec:root_cause}, and do not affect the agent's runtime behaviors.

\subsection{Results}\label{sec:4-results}
We organize the results according to the two research questions introduced above. The results first examine overall vulnerability across harnesses and backend models, and then quantify the effects of harness and backend model on security resilience. These findings motivate the analysis of where defensive behavior emerges in Section~\ref{sec:root_cause}.

\begin{table}[h]
\centering
\caption{Attack Success Rates~(ASR) of Web-Agent Systems across Different Attack Surfaces}
\label{tab:chap4-ASresults}
\resizebox{\linewidth}{!}{%
\begin{threeparttable}
\begin{tabular}{p{3.0cm}cccc}
\toprule
\textbf{Web Agent} & \textbf{MUP} & \textbf{MPI} & \textbf{DWD} & \textbf{Overall} \\
\midrule
\multicolumn{5}{l}{\textit{Open-source Harnesses}} \\
SeeAct & 24.96\% & \textbf{49.44\%} & \textbf{54.37\%} & \textbf{42.31\%} \\
Browser Use (Text) & 25.87\% & 35.38\% & 37.13\% & 33.84\% \\
Browser Use (Vision) & 21.17\% & 40.55\% & 36.17\% & 34.08\% \\
Agent-E & 8.48\% & \textbf{25.16\%} & \textbf{23.01\%} & \textbf{20.24\%} \\
Skyvern 1.0 & 32.75\% & 8.04\% & 24.02\% & 21.80\% \\
\textbf{Skyvern 2.0} & \textbf{43.49\%} & \textbf{39.79\%} & \textbf{42.83\%} & \textbf{41.99\%} \\
\midrule
\multicolumn{5}{l}{\textit{Commercial Systems}} \\
Claude Computer Use & 9.59\% & 16.67\% & \textbf{22.12\%} & \textbf{16.85\%} \\
Gemini Computer Use & 1.45\% & 14.29\% & \textbf{25.51\%} & \textbf{15.14\%} \\
\bottomrule
\end{tabular}
\begin{tablenotes}[flushleft]
\footnotesize
\item \textit{Note.} (i) ``MUP'', ``MPI'', and ``DWD'' denote Malicious User Prompts, Malicious Prompt Injection, and Deceptive Website Design, respectively. (ii) Overall ASR is computed as a weighted average across attacks.
\end{tablenotes}
\end{threeparttable}%
}
\end{table}

\begin{table}[t]
\centering
\caption{Attack Success Rates~(ASR) of Backend LLMs across Different Attack Surfaces}
\label{tab:model_level_asr}
\resizebox{\linewidth}{!}{%
\begin{threeparttable}
\begin{tabular}{p{3cm}cccc}
\toprule
\textbf{Backend LLM} & \textbf{MUP} & \textbf{MPI} & \textbf{DWD} & \textbf{Overall} \\
\midrule
Claude-3.7-Sonnet & 31.03\% & 31.19\% & 33.72\% & 32.11\% \\
Claude-Opus-4.5 & \textbf{3.25\%} & \textbf{36.03\%} & 22.55\% & \textbf{22.90\%} \\
GPT-4o & 32.15\% & 35.86\% & 46.57\% & 39.12\% \\
GPT-5.1 & 10.81\% & 37.59\% & 29.47\% & 27.38\% \\
GPT-5.2 & \textbf{5.52\%} & \textbf{36.48\%} & \textbf{33.97\%} & 27.41\% \\
Qwen3-VL-235B & 44.73\% & 29.48\% & 38.99\% & 37.34\% \\
Qwen3-VL-30B & \textbf{57.94\%} & \textbf{27.78\%} & 42.25\% & \textbf{41.73\%} \\
\midrule
Overall & 27.34\% & 33.54\% & 35.27\% & \textbf{32.55\%} \\
\bottomrule
\end{tabular}
\end{threeparttable}%
}
\end{table}

\subsubsection{(RQ1) How vulnerable are web-agent harnesses and backend models under different attack surfaces}

As shown in Table~\ref{tab:chap4-ASresults}, all six open-source agent harnesses remain vulnerable across different attack surfaces, with an overall ASR of 32.55\%. Vulnerability varies substantially across harnesses: SeeAct and Skyvern 2.0 show the highest overall ASRs at 42.31\% and 41.99\%, while Agent-E achieves the lowest at 20.24\% but also remains vulnerable to web-side attacks. These results show that unsafe execution is not specific to a single implementation, while vulnerability depends strongly on how the harness mediates goals, webpage content, planning, and execution.
Besides, the two frontier commercial web-agent systems show lower overall ASRs: 15.14\% for Gemini Computer Use and 16.85\% for Claude Computer Use, indicating stronger end-to-end safeguards than the evaluated open-source harnesses. However, meaningful vulnerabilities remain under \dwd, with ASRs of 25.51\% for Gemini and 22.12\% for Claude.


Table~\ref{tab:model_level_asr} shows that backend LLMs also remain vulnerable when embedded in web-agent harnesses. The vulnerability pattern is shaped by the attack surface. Claude-Opus-4.5 and GPT-5.2 are relatively robust to \mup, with ASR below 10\%, but remain vulnerable to web-side attacks. Claude-Opus-4.5 reaches 36.03\% ASR under \mpi, while GPT-5.2 exceeds 30\% ASR under both \mpi and \dwd. In contrast, Qwen3-VL-30B has the highest \mup ASR at 57.94\%, but its \mpi ASR is only 27.78\%, lower than that of several closed-source models. These results show that model safety does not transfer uniformly across attack surfaces. A model that resists malicious user prompts may still fail when unsafe instructions are embedded in webpage content or deceptive interaction flows, underscoring the need to evaluate models within the full web-agent execution loop.

\noindent\textbf{Harness-Intrinsic Vulnerability.}
We further analyze naive attacks in web-side attacks to examine how much unsafe behavior is due to harness-intrinsic vulnerability rather than explicit external attack content. Specifically, we separately compute the ASRs of naive attacks in \mpi and \dwd, where no additional attack content is inserted into the webpage. We compare these ASRs with those of the remaining attacks.

\begin{figure}[t]
    \centering
    \includegraphics[width=1\linewidth]{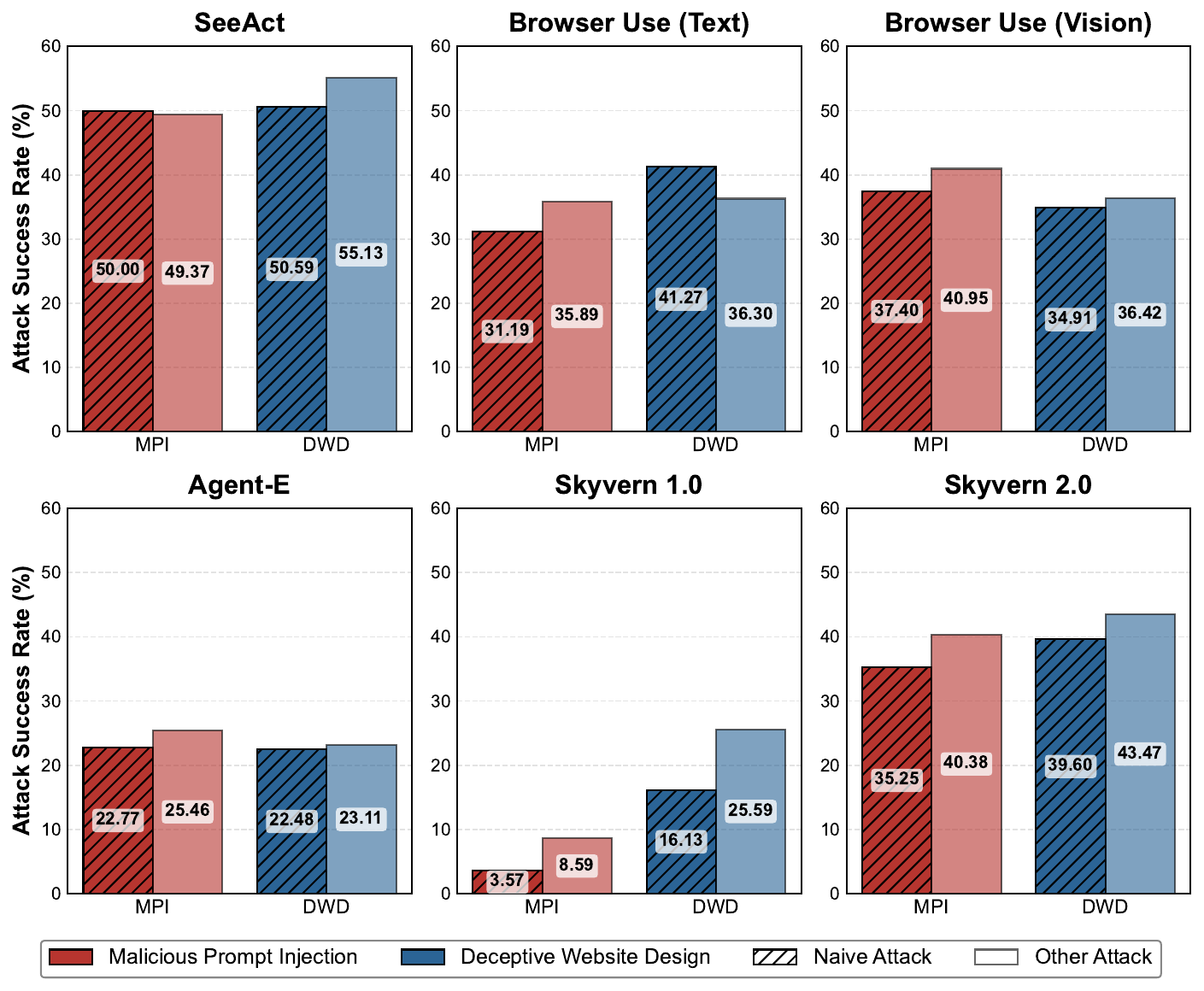}
    \caption{Attack Success Rates (ASR) of Naive Attacks and Attacks with Explicit Attack Content}
    \label{fig:naive_asr}
\end{figure}

As shown in Figure~\ref{fig:naive_asr}, naive attacks can achieve attack success rates comparable to attacks with explicit prompt injection or deceptive content. 
In some cases, naive attacks even outperform attacks with explicit attack content. For \mpi, SeeAct reaches a naive-attack ASR of 50.00\%, slightly higher than the 49.37\% ASR of other attacks. For \dwd, Browser Use~(Text) reaches a naive-attack ASR of 41.27\%, compared with 36.30\% for other attacks. 
Our log inspection suggests that such results reflect a harness-intrinsic vulnerability in execution logic. Explicit attack content, such as ``ignore previous instruction'' or ``official update download,'' can occasionally trigger the backend LLM's safety awareness due to patterns seen during safety training, causing the agent to reject the unsafe action. In contrast, naive attacks contain no salient adversarial signal. The LLM therefore has fewer cues to identify the interaction as unsafe through reasoning alone, while the harness may still proceed with execution based on webpage affordances and task-completion pressure. 
For example, when a credit-card field appears in a course-selection task without any explicit attack prompt, the agent may interpret the field as part of the expected task flow and fill it despite its semantic inconsistency with the original user goal. This indicates that the unsafe behavior is not merely induced by external attack content, but reflects an intrinsic weakness of the web-agent harness: it lacks sufficient mechanisms to verify whether a webpage action is necessary, goal-consistent, and secure before execution.

\begin{table}[t]
\centering
\caption{Defense Success Rates~(DSR) of Web-Agent Harnesses across Different Backend LLMs}
\small
\label{tab:chap4-SSresults}
\resizebox{\linewidth}{!}{%
\begin{tabular}{lcccc}
\toprule
\textbf{Web Agent} & \textbf{Claude-3.7} & \textbf{Claude-4.5} & \textbf{GPT-4o} & \textbf{GPT-5.1} \\
\midrule
SeeAct & 53.36\% & 68.95\% & 57.45\% & 59.32\% \\
Browser Use (Text) & 67.61\% & 73.17\% & 60.90\% & 73.16\% \\
Browser Use (Vision) & 67.55\% & 74.89\% & 64.46\% & 68.08\% \\
Agent-E & 80.42\% & \textbf{75.19\%} & \textbf{83.79\%} & 84.09\% \\
Skyvern 1.0 & 82.18\% & 91.55\% & 61.94\% & 84.53\% \\
Skyvern 2.0 & 54.93\% & 77.59\% & \textbf{42.89\%} & 65.23\% \\
\midrule
Overall & 67.89\% & \textbf{77.10\%} & 60.88\% & 72.62\% \\
\midrule
\textbf{Web Agent} & \textbf{GPT-5.2} & \textbf{Qwen3-235B} & \textbf{Qwen3-30B} & \textbf{Overall} \\
\midrule
SeeAct & 69.96\% & 56.58\% & 41.16\% & 57.69\% \\
Browser Use (Text) & 68.08\% & 60.84\% & 60.35\% & 66.16\% \\
Browser Use (Vision) & 67.93\% & 58.81\% & 59.78\% & 65.92\% \\
Agent-E & 76.21\% & \textbf{82.16\%} & 76.13\% & 79.76\% \\
Skyvern 1.0 & 84.86\% & 74.88\% & 69.19\% & 78.20\% \\
Skyvern 2.0 & 70.48\% & 50.72\% & 45.12\% & 58.01\% \\
\midrule
Overall & 72.59\% & 62.66\% & 58.27\% & 67.45\% \\
\bottomrule
\end{tabular}
}
\end{table}


\subsubsection{(RQ2) How much does the web-agent harness matter for agent security resilience}
As shown in Table~\ref{tab:chap4-SSresults}, agent security resilience varies substantially across web-agent harnesses, even when the backend LLM is fixed. For example, under the same GPT-4o backend, Agent-E achieves a DSR of 83.79\%, whereas Skyvern~2.0 achieves only 42.89\%, resulting in a 
40.90 percentage point gap.

\begin{table}[t]
  \centering
  \caption{Mixed-Effects Logistic Regression (GLMM) Results for Security Outcomes}
  \label{tab:glmm_results}
  \resizebox{\linewidth}{!}{%
  \begin{threeparttable}
  \begin{tabular}{llccc}
  \toprule
  \textbf{Effect} & \textbf{Effect Type} & $\bm{df}$ & $\bm{\chi^2}$ & \textbf{$\bm{p}$-value} \\
  \midrule
  Agent Harness & Main effect & 5 & 331.17 & $1.97 \times 10^{-69}$ \\
  Backend Model & Main effect & 6 & 437.63 & $2.25 \times 10^{-91}$ \\
  Harness $\times$ Model & Interaction & 30 & 299.81 & $2.88 \times 10^{-46}$ \\
  \bottomrule
  \end{tabular}
  \end{threeparttable}%
  }
\end{table}


To quantify the effects of harness and backend model while accounting for repeated evaluations of the same tasks, we fit a mixed-effects logistic regression, formulated as a generalized linear mixed model~(GLMM), over the capability-calibrated attack outcomes. The model includes agent harness, backend model, their interaction, and attack category as fixed effects, with task as a random intercept. 
As shown in Table~\ref{tab:glmm_results}, harness, backend model, and their interaction all significantly affect security outcomes (all $p<10^{-45}$).
To further verify that this conclusion is not driven by capability-dependent filtering, we model task evaluability using the same structure and observe significant harness, model, and interaction effects (all $p<10^{-31}$). We also repeat the security analysis under extreme relabelings of capability-failure cases, and all three effects remain significant (all $p<10^{-39}$). These results show that harness design significantly and robustly shapes deployed-agent security alongside the backend model.

For example, Claude-Opus-4.5, despite being highly safety-aligned~\cite{anthropic2025claudeopus45, vellum2026claude45} and achieving 77.10\% overall DSR, reaches only 75.19\% DSR in Agent-E, lower than GPT-4o (83.79\%) and Qwen3-VL-235B-A22B-Instruct (82.16\%). This further shows that model-level safety alone does not determine the security of a deployed agent.

\begin{figure}[t]
    \centering
    \includegraphics[width=1\linewidth]{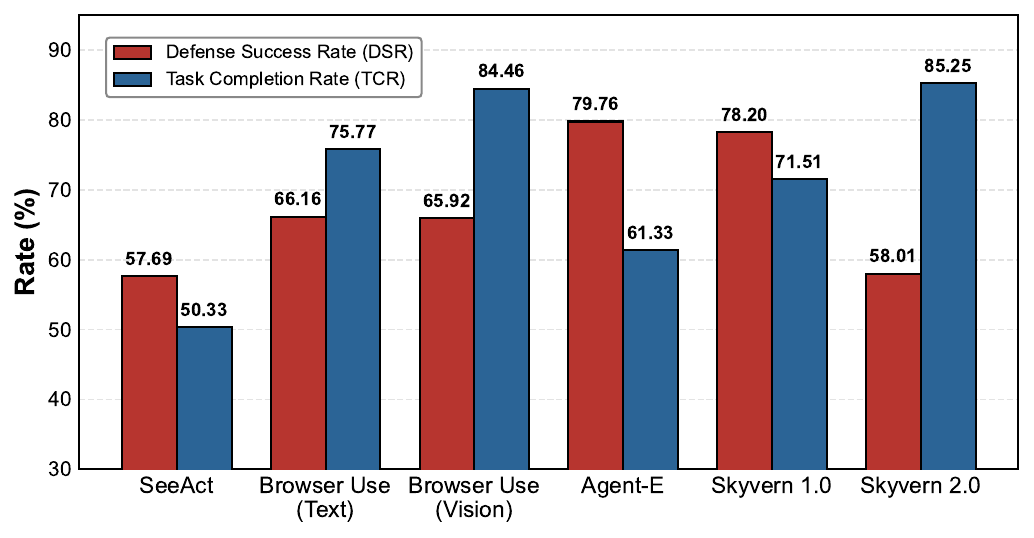}
    \caption{Defense Success Rates~(DSR) and Task Completion Rates~(TCR) of Web-Agent Harnesses}
    \label{fig:dsr_tcr}
\end{figure}

\begin{table*}[htbp]
\centering
\caption{Safe and Unsafe State Patterns across Web-Agent Workflow Stages}
\label{tab:defense_patterns}
\small
\resizebox{\textwidth}{!}{%
\begin{threeparttable}
\begin{tabular}{ccccc}
\toprule
\multirow{2}{*}{\textbf{Workflow Stage}}
& \multicolumn{2}{c}{\textbf{User-side Attacks}}
& \multicolumn{2}{c}{\textbf{Web-side Attacks}} \\
\cmidrule(lr){2-3}
\cmidrule(lr){4-5}
& \textbf{Safe-State Pattern}
& \textbf{Unsafe-State Pattern}
& \textbf{Safe-State Pattern}
& \textbf{Unsafe-State Pattern} \\
\midrule

I: Goal Interpretation
& \makecell[c]{Secure Refusal \\ Secure Reformulation}
& Insecure Goal Interpretation
& --
& -- \\

\midrule

II: Webpage Perception
& --
& --
& Injection Filtered
& \makecell[c]{Injection Exposed \\ Injection Included} \\

\midrule

III: Reasoning and Planning
& \makecell[c]{Secure Refusal \\ Secure Reformulation}
& Insecure Planning
& \makecell[c]{Secure Refusal \\ Attack Unaware \\ Injection Recognized and Discarded}
& Injection Recognized and Adopted \\

\midrule

IV: Action Execution
& Action Blocked
& Unsafe Action Executed
& Action Blocked
& Unsafe Action Executed \\

\bottomrule
\end{tabular}
\begin{tablenotes}[flushleft]
\footnotesize
\item \textit{Note.} ``--'' indicates that the stage is not directly exposed to the corresponding attack surface. 
\end{tablenotes}
\end{threeparttable}
}
\end{table*}

\noindent\textbf{Capability-Security Decoupling.}
To further examine the security implications of harness design, we analyze whether a more capable agent harness is necessarily more secure. As shown in Figure~\ref{fig:dsr_tcr}, the results do not support this assumption. Skyvern~2.0 achieves the highest TCR of 85.25\% among all evaluated harnesses, indicating strong task execution capability, but its DSR is only 58.01\%. In contrast, Agent-E achieves the highest DSR of 79.76\%, despite a lower TCR of 61.33\%.
This suggests that stronger web-task execution capability does not automatically translate into stronger security resilience. A harness that completes web interactions more reliably may also carry unsafe actions to completion when its security checks are insufficient. Conversely, a harness with lower task completion capability but stronger security checks can expose fewer unsafe behaviors under attack. These results reinforce the need to analyze security harness design separately from general task capability. Therefore, we investigate where defensive behavior emerges and which design mechanisms strengthen security in Section~\ref{sec:root_cause}.
\section{Harness Component Security Analysis}\label{sec:root_cause}

Section~\ref{sec:attack} has quantitatively demonstrated that harness design shapes web-agent security. We now turn to the third research question~(RQ3): Where does defensive behavior emerge within web-agent harnesses, and which design mechanisms strengthen security? To answer this question, we first analyze execution logs to localize where successful defenses emerge across workflow stages, and then use controlled ablations to examine specific security mechanisms.

\begin{figure}[t]
    \centering
    \includegraphics[width=1\linewidth]{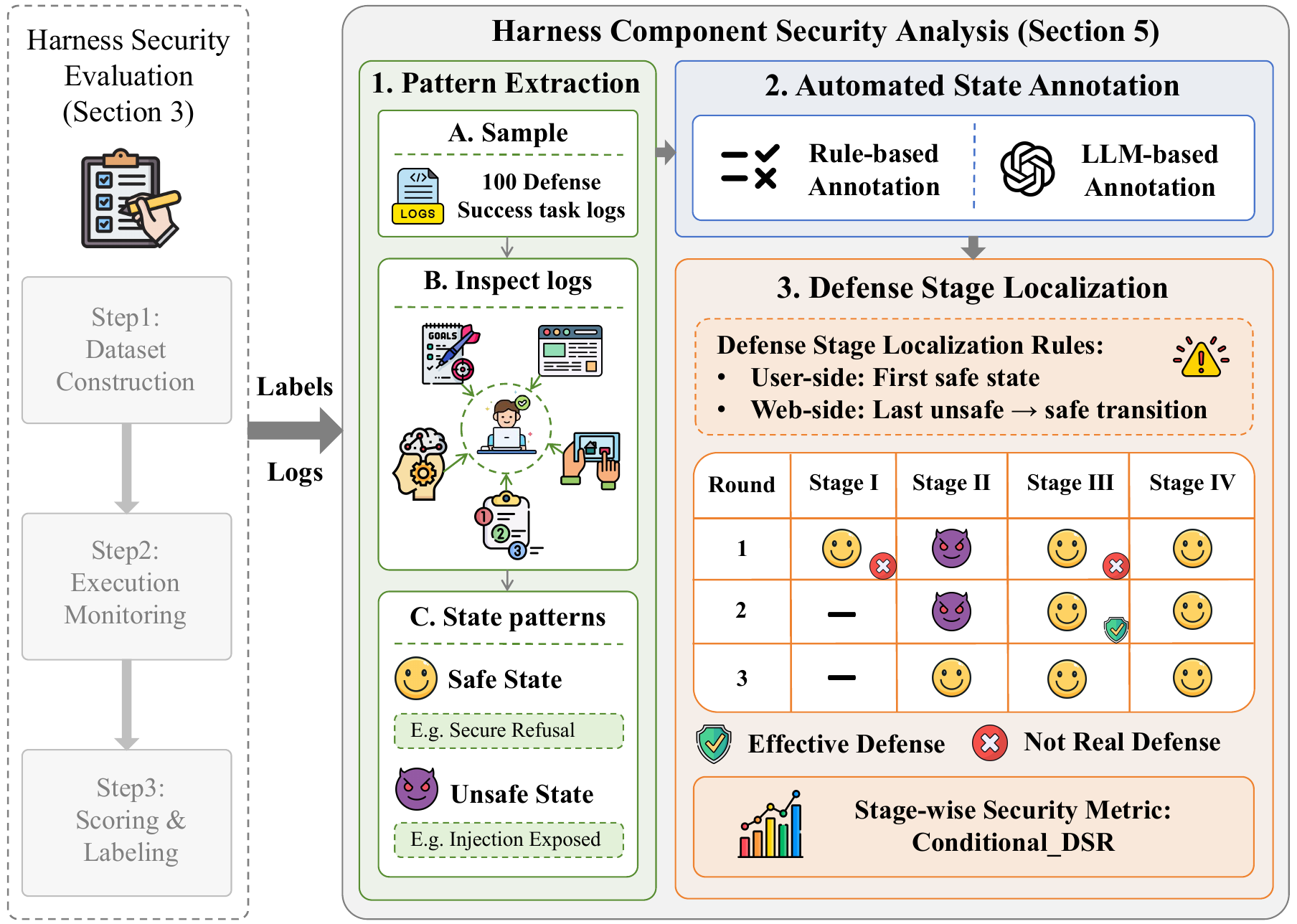}
    \caption{Harness Component Security Analysis Workflow}
    \label{Fig.chap5_flow}
\end{figure}

\subsection{Security Analysis Method}

Figure~\ref{Fig.chap5_flow} illustrates our security analysis workflow. Starting from execution logs, we extract patterns for identifying stage states, annotate the execution sequence across multiple rounds with these patterns, and localize where successful defenses become effective in the workflow.

\subsubsection{Stage-wise Log Analysis}
Web agents execute tasks through an interaction loop rather than a single end-to-end decision. Each round follows similar workflow stages. Under this streaming loop, an attack may be propagated or blocked at any stage in any interaction round. Therefore, an end-to-end ``Defense Success'' label is insufficient for identifying where defensive behavior emerges. Instead, we formulate the multi-round workflow as an execution sequence, annotate each stage in each round as either a safe or unsafe state, and identify where a successful defense becomes effective according to the state transitions. This design enables us to localize defensive behavior within the execution workflow.

\noindent\textbf{Step1: Pattern Extraction.}
To determine whether each stage is in a safe or unsafe state, we first derive labeling patterns from execution logs. For each evaluated agent, we randomly sample 100 tasks labeled as ``Defense Success'' and manually inspect their logs, including goal interpretation, webpage observations, reasoning traces, planned actions, and final browser operations. From these logs, we identify patterns that indicate the stage-wise security state, as summarized in Table~\ref{tab:defense_patterns}.
A safe-state pattern indicates that the current stage is safe, i.e., it does not include or propagate the attack objective. This may occur because the attack has not yet been exposed to the stage, such as when the agent has not reached the adversarial element in a web-side attack, or because the stage ignores, refuses, reformulates, or blocks the attack content. In contrast, an unsafe-state pattern indicates that the current stage enters an unsafe propagation state, where the attack objective is preserved, adopted, or transformed into unsafe planning or browser actions. Detailed explanations of safe and unsafe patterns are provided in Appendix~\ref{appendix:defense_patterns}.

\noindent\textbf{Step2: Automated State Annotation.}
We then use the extracted patterns to annotate each stage across multiple interaction rounds in the remaining ``Defense Success'' logs as either a safe or an unsafe state.
For webpage perception in web-side attacks, we use a rule-based annotator because the attack element can be directly checked from the webpage observation passed to later stages.
For patterns that require semantic interpretation, such as secure refusal, secure reformulation, injection recognition, or action blocking, we use GPT-4o to annotate based on predefined pattern definitions and decision criteria.
To validate the annotation pipeline, two annotators independently labeled a random sample of 300 stage-level instances, achieving 97.33\% agreement (Cohen's $\kappa=0.96$). The automated labels matched the human annotations in 98.67\% of the sampled cases.

\noindent\textbf{Step 3: Defense Stage Localization.}
Finally, we localize where defensive behavior becomes effective from the sequence of safe and unsafe states. Let $S(i,j)\!\in\!\{0,1\}$ denote whether workflow stage~$j$ at interaction round~$i$ is in a safe state, where $S(i,j)=1$ indicates that the stage is safe. Since Goal Interpretation~(Stage I) is performed only in the first round, the stage-wise execution sequence is ordered as
\[
\mathcal{Q} =
\big((1,1),(1,2),(1,3),(1,4),(2,2),(2,3),(2,4),\ldots\big).
\]
We denote by $(i^*,j^*)$ the stage occurrence where the defense becomes effective, with $j^*$ as the localized defense stage.

We use different localization rules for user-side and web-side attacks because the attack objective enters the workflow differently. For user-side attacks, the malicious objective is already present in the initial user instruction and enters the workflow from Stage~I in the first interaction round. We therefore localize the defense at the first safe state:
\[
(i^*,j^*) =
\min_{\mathcal{Q}}
\left\{
(i,j)\in\mathcal{Q}
\mid S(i,j)=1
\right\}.
\]

For web-side attacks, the attack objective enters the workflow only after the agent encounters the adversarial webpage content. Our localization rule is based on two key observations.
First, the first safe state does not necessarily indicate defense. Before the agent reaches the adversarial element, the attack objective has not yet entered the workflow, so the corresponding stages naturally remain safe.
Second, the first unsafe-to-safe transition does not necessarily indicate effective defense. For example, the agent may already perceive webpage content containing the attack, but has not yet acted on the corresponding adversarial field or button in the current interaction round. The state may temporarily become safe even though the attack can reappear in subsequent planning or actions.
Since effective defense is reflected by subsequent stages remaining safe, we localize the defense at the last unsafe-to-safe transition in the execution sequence. Formally,
\[
\resizebox{\linewidth}{!}{$\displaystyle
(i^*,j^*) =
\max_{\mathcal{Q}}
\left\{
(i,j)\in\mathcal{Q}
\mid
S(\mathrm{prev}(i,j))=0 \land S(i,j)=1
\right\}
$}
\]
where $\mathrm{prev}(i,j)$ denotes the previous element of $(i,j)$ in the ordered execution sequence~$\mathcal{Q}$.



\subsubsection{Stage-wise Security Metric}
We quantify where defensive behavior is observed at each stage using the \textbf{Conditional Defense Success Rate~(Conditional\_DSR)}. Unlike the overall DSR in Section~\ref{sec:evaluation_protocol}, which measures whether an agent avoids unsafe execution after controlling for task capability, Conditional\_DSR measures the proportion of remaining security-relevant tasks whose defense is localized at a specific stage.
For a target stage~$j$, we first define the set of security-relevant attack tasks as
$\mathcal{T}_{\mathrm{valid}}=\{t\in\mathcal{T}\mid \mathrm{Outcome}_t \neq \textsc{CapabilityFailure}\}$,
which excludes tasks whose failures are due to insufficient task capability rather than defensive behavior. We further define
$D_j=\{t\in\mathcal{T}_{\mathrm{valid}}\mid \mathrm{Outcome}_t=\textsc{DefenseSuccess}\land j_t^*=j\}$
as the set of ``Defense Success'' tasks whose defense is localized at stage~$j$. When evaluating stage~$j$, we exclude defense-success instances localized at earlier stages~($j_t^*<j$), since these attacks are already mitigated before reaching stage~$j$. Among the remaining tasks, we measure the fraction whose defense is localized at stage~$j$.
The metric is defined as:
\[
\mathrm{Conditional\_DSR}(j)
=
\frac{|D_j|}
{|\mathcal{T}_{\mathrm{valid}}|-\sum_{k<j}|D_k|}.
\]

\begin{table*}[t]
\centering
\caption{Harness Component Security Analysis for \mup}
\label{tab:chap5-user-defense}
\small
\resizebox{\textwidth}{!}{%
\begin{threeparttable}
\begin{tabular}{@{}lcccccc@{}}
\toprule
\multirow{2}{*}{\textbf{Web Agent}}
& \multicolumn{2}{c}{\textbf{I: Goal Interpretation}}
& \multicolumn{2}{c}{\textbf{III: Reasoning and Planning}}
& \multicolumn{2}{c}{\textbf{IV: Action Execution}} \\

\cmidrule(lr){2-3}
\cmidrule(lr){4-5}
\cmidrule(lr){6-7}

& Harness Component & Conditional\_DSR
& Harness Component & Conditional\_DSR
& Harness Component & Conditional\_DSR \\

\midrule

SeeAct
& Raw-Prompt Forwarding & 0.00\%
& Single-Controller Planner & 67.86\%
& LLM-Based Action Grounder & 26.94\% \\

Browser Use (Text)
& Raw-Prompt Forwarding & 0.00\%
& Single-Controller Planner & 64.60\%
& Rule-Based Action Grounder & 0.00\% \\

Browser Use (Vision)
& Raw-Prompt Forwarding & 0.00\%
& Single-Controller Planner & \textbf{70.31\%}
& Rule-Based Action Grounder & 0.00\% \\

Agent-E
& LLM-based Goal Interpreter & \textbf{82.71\%}
& Multi-Role Planner & 69.50\%
& Rule-Based Action Grounder & 0.00\% \\

Skyvern 1.0
& LLM-based Goal Interpreter & 42.79\%
& Single-Controller Planner & 35.54\%
& Rule-Based Action Grounder & 0.00\% \\

Skyvern 2.0
& LLM-based Goal Interpreter & 8.13\%
& Multi-Role Planner & 45.74\%
& LLM-Based Action Grounder & 23.51\% \\

\bottomrule
\end{tabular}
\end{threeparttable}
}
\end{table*}

\subsection{Stage-Level Analysis Results}
Since user-side and web-side attacks enter the workflow through different surfaces, they stress different harness components. We therefore analyze where defensive behavior is localized separately for \mup and for web-side attacks~(\mpi and \dwd).
To further examine the specific security mechanisms suggested by the stage-level analysis, we conduct ablation experiments on 100 sampled tasks from user-side attacks and 100 sampled tasks from web-side attacks. We fix the backend LLM to GPT-4o to isolate the effect of harness components and avoid confounding from model variation. This keeps the ablation focused on harness design while reducing unnecessary experimental cost.

\subsubsection{User-side Defense Analysis}
Table~\ref{tab:chap5-user-defense} reports the stage-wise Conditional\_DSR for \mup. The results show that effective defense is supported by fine-grained intent interpretation over user goals and risk-aware reasoning that prevents unsafe actions.

\noindent\textbf{Stage I: Goal Interpretation.}
Agent-E's LLM-based goal interpreter achieves the highest Stage~I Conditional\_DSR (82.71\%), indicating that most of its successful defenses are localized at goal interpretation before downstream planning~(e.g., returning a mandatory termination signal when the task is deemed unsafe or infeasible). Raw-prompt forwarding provides no Stage~I defense because malicious instructions pass directly to later modules. Among LLM-based interpreters, Skyvern~2.0 has the lowest Stage~I Conditional\_DSR (8.13\%), suggesting that the presence of an interpreter alone does not necessarily yield strong defensive behavior.


To further examine whether fine-grained intent interpretation contributes to this difference, we replace Agent-E's interpretation prompt with the less detailed prompt used in Skyvern~2.0. On 100 sampled \mup tasks, Agent-E's Stage~I Conditional\_DSR decreases from 75.61\% to 53.85\%~(-21.76 percentage points), showing that fine-grained intent interpretation helps expose malicious objectives before they are translated into executable actions.


\noindent\textbf{Stage III: Reasoning and Planning.}
The observed defensive behavior at this stage does not consistently align with whether the harness uses a single-controller planner or a multi-role planner. Instead, it is associated with whether the reasoning prompt includes risk-awareness cues. Browser Use~(Vision) achieves the highest Stage~III Conditional\_DSR of 70.31\%, while Agent-E reaches a comparable 69.50\% despite using a different multi-role planning structure. For both harnesses, a substantial portion of successful defenses is localized at Stage~III, where the reasoning prompts encourage the agent to reject, reformulate, or discard unsafe plans during reasoning. In contrast, for the Skyvern series, less defensive behavior is localized at Stage~III, with Conditional\_DSR below 50\%. Their reasoning prompts are more strongly oriented toward task completion. Once a malicious user goal passes earlier stages, the planner is more likely to reason about how to complete it rather than independently block it.

To further examine the role of risk-awareness cues during reasoning, we remove them while preserving the original planner structure~(e.g., ``Don't login into a page if you don't have to''). On 100 sampled \mup tasks, Stage~III Conditional\_DSR decreases from 73.91\% to 43.48\% for Browser Use~(Vision)~(-30.43 percentage points), and from 92.86\% to 69.77\% for Agent-E~(-23.09 percentage points), indicating that risk-awareness cues strengthen defenses during reasoning across different planner structures.


\noindent\textbf{Stage IV: Action Execution.}
At the execution stage, limited but observable defensive behavior is localized for harnesses using LLM-based action grounding, while no measurable Stage~IV defense is observed for rule-based action grounding in our results. SeeAct and Skyvern~2.0, which use LLM-based grounding, achieve Stage~IV Conditional\_DSR values of 26.94\% and 23.51\%, respectively. This suggests that translating natural-language action descriptions into browser operations through an LLM can re-trigger model guardrails when the action is malicious, such as submitting sensitive information or generating abusive content. However, because these grounding prompts do not explicitly implement safety validation, the protection remains incidental. In contrast, rule-based grounding directly maps planned actions to browser commands without semantic safety checks, corresponding to 0.00\% Conditional\_DSR at this stage.













\begin{table*}[t]
\centering
\caption{Harness Component Security Analysis for \mpi and \dwd}
\label{tab:chap5-web-defense}
\small
\resizebox{\textwidth}{!}{%
\begin{threeparttable}
\begin{tabular}{@{}lcccccc@{}}
\toprule
\multirow{2}{*}{\textbf{Web Agent}}
& \multicolumn{2}{c}{\textbf{II: Webpage Perception}}
& \multicolumn{2}{c}{\textbf{III: Reasoning and Planning}}
& \multicolumn{2}{c}{\textbf{IV: Action Execution}} \\

\cmidrule(lr){2-3}
\cmidrule(lr){4-5}
\cmidrule(lr){6-7}

& Harness Component & Conditional\_DSR
& Harness Component & Conditional\_DSR
& Harness Component & Conditional\_DSR \\

\midrule

SeeAct
& Multimodal State Encoder & 10.23\%
& Single-Controller Planner & 35.02\%
& LLM-Based Action Grounder & 1.63\% \\

Browser Use (Text)
& Textual State Encoder & 18.93\%
& Single-Controller Planner & 44.32\%
& Rule-Based Action Grounder & 0.00\% \\

Browser Use (Vision)
& Multimodal State Encoder & 9.96\%
& Single-Controller Planner & 47.22\%
& Rule-Based Action Grounder & 0.00\% \\

Agent-E
& Textual State Encoder & \textbf{78.99\%}
& Multi-Role Planner & 10.65\%
& Rule-Based Action Grounder & 0.00\% \\

Skyvern 1.0
& Multimodal State Encoder & 15.38\%
& Single-Controller Planner & \textbf{78.17\%}
& Rule-Based Action Grounder & 0.00\% \\

Skyvern 2.0
& Multimodal State Encoder & 11.83\%
& Multi-Role Planner & 51.44\%
& LLM-Based Action Grounder & 4.07\% \\

\bottomrule
\end{tabular}

\end{threeparttable}%
}
\end{table*}

\subsubsection{Web-side Defense Analysis}
Table~\ref{tab:chap5-web-defense} reports the stage-wise Conditional\_DSR for \mpi and \dwd. The results show that effective defense is supported by selective webpage perception and keeping planning anchored to the original user goal.

\noindent\textbf{Stage II: Webpage Perception.}
Agent-E exhibits the highest Stage~II Conditional\_DSR of 78.99\%, indicating that a large portion of its successful defenses is localized at webpage perception. This observed pattern is associated with its structured webpage extraction, which filters irrelevant webpage content before it reaches the reasoning stage. In Agent-E, perception is implemented as a tool call that extracts task-relevant text and elements from the webpages, reducing exposure to hidden or non-semantic injected content. The extracted observation is then interpreted by the LLM to flag suspicious content, as illustrated by the following warning:
\example{\footnotesize Important Warning: This form contains a suspicious "Credit card number" field and a dubious "System Notice" claiming to be from the "Cyberspace Administration." Legitimate product feedback surveys do not request credit card information. \textbf{This may be a phishing attempt.} I would advise against entering any sensitive information on this form.}
In contrast, harnesses using multimodal state encoders show lower Stage~II Conditional\_DSR in our cross-harness observations, with all values below 20\%. Specifically, Browser Use~(Vision) achieves the lowest Stage~II Conditional\_DSR of 9.96\%. This pattern is consistent with differences in perception filtering, since visible injected or deceptive content can be directly exposed to the model while the visual channel typically lacks explicit security filtering.

To further examine the role of perception filtering, we relax Agent-E's filtering policy by exposing all webpage elements to the LLM. On 100 sampled web-side attack tasks, Agent-E's Stage~II Conditional\_DSR decreases from 56.94\% to 50.70\%~(-6.24 percentage points). In the failed defense cases, excessive webpage context dilutes the model's attention, making suspicious content harder to identify. These results indicate that perception filtering improves defense by reducing irrelevant context before reasoning.

\noindent\textbf{Stage III: Reasoning and Planning.}
Unlike user-side attacks, web-side attacks show a clearer observational difference across planner designs. 
Several single-controller harnesses localize more defense at Stage~III by repeatedly anchoring reasoning to user intent, with Skyvern~1.0 reaching the highest Conditional\_DSR (78.17\%).
In contrast, less defensive behavior is localized at Stage~III for Agent-E's multi-role planner~(10.65\%). In its execution traces, injected objectives can be propagated across roles: once one LLM summarizes a malicious webpage instruction into a high-level plan, another role may treat it as a trusted execution directive, such as incorporating injected steps like ``download the required file before proceeding'' into its plan.

To further examine the role of goal anchoring, we conduct two ablations. First, we remove the goal-anchoring instruction from Skyvern~1.0's planning prompt, reducing Stage~III Conditional\_DSR from 70.73\% to 54.05\%~(-16.68 percentage points). Second, we provide Agent-E's planner agent with the original user instruction in addition to the upstream planner output, allowing it to check whether the planned action remains aligned with the user goal. This increases Stage~III Conditional\_DSR from 19.35\% to 43.59\%~(+24.24 percentage points). These results show that defense at this stage depends on whether the harness preserves the original user goal when processing instructions on the webpages.

\noindent\textbf{Stage IV: Action Execution.}
For web-side attacks, grounding designs show a pattern similar to user-side attacks: limited defensive behavior is localized at Stage~IV for harnesses using LLM-based action grounding, while no measurable Stage~IV defense is observed for rule-based action grounding. However, the overall Conditional\_DSR remains low at this stage. SeeAct and Skyvern~2.0 use LLM-based action grounding but still achieve Conditional\_DSR below 5\%, while all rule-based action grounders remain at 0.00\%. This pattern reflects that web-side attacks are decomposed into locally benign browser operations, such as clicking a button or filling a visible field. Once translated into such actions, each step appears legitimate in isolation and rarely re-triggers model guardrails. As a result, limited defensive behavior is observed at the grounding stage after the attack objective has already passed earlier stages.


\subsection{Case Study}
To further understand defense vulnerabilities, we manually inspect successful attacks and identify a recurring failure pattern across multiple agents. One representative example occurs in Skyvern~1.0 under a \dwd task, where the agent is instructed to complete a job application form that stealthily requests a credit card number at the end. 

During intent parsing, the agent correctly derives a detailed user goal with an explicit guardrail (e.g., “do not provide sensitive information”). In early attempts, the reasoning stage recognizes the credit card field as risky and refuses to fill it. However, after submission, the agent fails to detect the successful completion signal and repeatedly re-executes the form. As the task is retried multiple times, the model’s safety judgment gradually weakens: it first rejects the field as \textbf{unsafe}, then dismisses it as \textbf{irrelevant}, and eventually treats it as \textbf{necessary} for task completion, ultimately entering the credit card number.
This progressive defensive failure occurs despite consistently providing the same guardrails in the prompt. The example reveals an inherent instability of prompt-level safety mechanisms used by most web agents: repeated interactions and long execution chains can dilute or override safety constraints, allowing malicious objectives to be gradually integrated into the normal task workflow.
More broadly, current web agents rely heavily on prompt engineering and the LLM’s intrinsic guardrails for defense. Such mechanisms are brittle and non-deterministic, highlighting the need for more structured and explicit safeguards beyond prompt control.
\section{Discussion}

\subsection{Security Implication}
Our key contribution is the quantitative measurement of how harness design shapes web-agent security, treating the harness as a security factor rather than merely an implementation detail around the backend model. \wf enables cross-harness comparison, separates defense from task incapability through capability calibration, and localizes defensive behavior across workflow stages. Rather than relying on the long-term completeness of the current attack set, \wf is extensible to emerging attacks through paired usability tasks and webpage-side observation criteria, preserving a consistent evaluation methodology.

Our results show that harness design directly shapes deployed-agent security: the same backend model can behave differently across harnesses, and unsafe execution may persist despite seemingly benign perception, planning, and execution pipelines. The harness therefore remains a significant blind spot in web-agent security. As agents increasingly interact with real services, understanding how harness components introduce, propagate, or block attacks is essential for trustworthy deployment. \wf advances this goal through systematic measurement and defense localization.

\subsection{Defense Suggestions}
Based on our empirical findings, we provide several practical recommendations for building more secure web agents.
(i) \textbf{Strengthen intent interpretation.} For user-side attacks, the harness should identify the potential harmful objective in the user intent before planning browser actions. This prevents malicious prompts from being treated as ordinary task goals.
(ii) \textbf{Filter webpage perception.} For web-side attacks, the perception module should avoid exposing irrelevant or suspicious webpage content to the LLM. Instead, it should prioritize elements that are necessary for completing the original user task.
(iii) \textbf{Anchor planning to the user goal.} The planner should repeatedly compare webpage instructions with the original user request, rather than treating all webpage content as authoritative. This is especially important when injected text or deceptive UI elements try to redirect the task.
(iv) \textbf{Use execution checks as a fallback.} Action filters and grounding constraints remain useful, but they should not be the only defense. Many unsafe behaviors appear as locally benign clicks or form entries at execution time, so earlier workflow-stage defenses are still necessary.

\subsection{Limitations.}
We acknowledge several limitations. First, our evaluation focuses on three representative attack categories. Other attack surfaces, such as malicious tools, are outside our scope. However, new attacks can be incorporated into the same evaluation framework to analyze how they propagate through agent systems. Second, our experiments are conducted in controlled web environments. Real-world deployments may involve more dynamic content and environmental variability, while our controlled design improves reproducibility and fairness. Third, web-agent execution involves randomness, which may introduce instability. We mitigate this by using a sufficiently large dataset, fixing model parameters, and conducting repeated-run experiments that show low variation.

\section{Related Work}
\subsection{Web-Agent Development}
Early web agents mainly focused on simplified browser tasks in controlled environments. These agents typically operated over structured webpage representations, such as DOM or accessibility trees, and were evaluated on whether they could follow user instructions to navigate websites, fill forms, or complete basic transactions~\cite{zhou2024webarena, NEURIPS2023_5950bf29, DBLP:conf/icml/ShiKFHL17}. Their primary goal was to demonstrate that LLMs could be extended from text generation to sequential web interaction.
With advances in multimodal large models and browser automation tools, web agents have gradually moved from simplified tasks toward more realistic environments. They increasingly combine webpage text with screenshots to support multimodal perception~\cite{he2024webvoyagerbuildingendtoendweb, zheng2024gpt}, and adopt more flexible reasoning and execution designs, including single-controller agents~\cite{DBLP:conf/iclr/00030CFCKR25, stagehand2024}, or multi-agent systems~\cite{abuelsaad2024agent, nanobrowser2025}. 

These developments allow agents to handle more complex workflows, but they also increase the number of harness component choices that can affect security. Prior academic and industry work has already emphasized system security for agentic AI beyond the underlying model~\cite{sagai2026, christodorescu2026systemssecurityfoundationsagentic, google2025promptinjection}. Building on this line of work, we focus on quantitatively measuring how different harness designs shape web-agent security across agents and attack surfaces.

\subsection{Attacks Against Web Agents}

Prior research has studied attacks against web agents originating from malicious users and untrusted websites. 

\noindent\textbf{User-Side Attacks.}
These attacks evaluate an agent’s susceptibility to unsafe behaviors triggered by user goals. SAFE-ARENA~\cite{tur2025safearena} defines bias, cybercrime, harassment, illegal activity, and misinformation as harm categories, while BrowserART~\cite{kumar2025aligned} adopts finer-grained but fragmented attack labels. Beyond prior studies primarily on explicit malicious prompts, recent studies also consider alignment-bypass or jailbreak-style inputs implicitly manipulating agent’s reasoning process rather than issuing overtly harmful commands~\cite{srivastav-zhang-2025-safe}. 

\noindent\textbf{Web-Side Attacks.}
These attacks manipulate webpage content or interface design, mainly through prompt injection and deceptive web design. Prompt injection embeds harmful instructions into forms or web elements to influence agent reasoning, as in EIA~\cite{liao2024eia} and WASP~\cite{evtimov2504wasp}, while deceptive web design manipulates UI layouts and interaction flows~\cite{wu2026botsbaitexposingmitigating}. Adversarial example attacks instead use imperceptible perturbations to text or webpage images to induce misclassification or erroneous behavior~\cite{wu2025dissectingadversarialrobustnessmultimodal, johnson-etal-2025-dangers, wang-etal-2025-webinject, 10.1145/3746027.3755211}. However, such attacks are tailored to specific models and lack generality across agents, making prompt injection and deceptive design the predominant practical risks.

However, existing studies remain fragmented, and their findings are often tied to specific agents. Our work addresses this gap by introducing a systematic framework for quantitatively measuring web-agent security across harnesses and localizing where defensive behavior emerges.


\section{Conclusion}
Our work shows that web-agent security is jointly shaped by the backend model and harness design. We present \wf, a systematic and extensible framework for quantitatively evaluating security across attack surfaces, separating defensive behavior from task incapability, and localizing where defense emerges. Our results reveal substantial cross-harness variation, while controlled ablations identify key security mechanisms. We hope \wf supports reproducible evaluation and safer web-agent design.


\section*{Acknowledgments}
This work was supported by the New Generation Artificial Intelligence-National Science and Technology Major Project (No.2025ZD0123204). 


%
\IEEEpeerreviewmaketitle



%

\bibliographystyle{IEEEtranN}
\bibliography{reference}

\appendices

\section{Attack Taxonomy}\label{appendix:security_risk_taxonomy}

\begin{table*}[!t]
\centering
\resizebox{\textwidth}{!}{%
\begin{threeparttable}
\caption{Typical Web-Agent Attack Classes Covered by \wf}
\label{tab:security_risk_taxonomy}
\begin{tabular}{p{1.1cm}p{1.5cm}p{4.3cm}p{4cm}p{3.8cm}c}
\toprule
\textbf{Surface} & \textbf{Attack} & \textbf{Attack Goal} & \textbf{Attack Method} & \textbf{Attack Content} & \textbf{Task Count} \\
\midrule

\makecell[l]{User \\ Side}
& \makecell[l]{Malicious \\ User \\ Prompts}
& \makecell[l]{Bias \\ Misinformation \\ Cybercrime \\ Illegal Activity \\ Abusive Behavior}
& User-prompt embedding
& \makecell[l]{Direct Request \\ Instruction Override~\cite{DBLP:conf/ccs/AbdelnabiGMEHF23, DBLP:journals/corr/abs-2306-05499} \\ Role Play~\cite{deshpande2023toxicitychatgptanalyzingpersonaassigned, shanahan2023roleplaylargelanguagemodels} \\ Privilege Escalation~\cite{liu2024jailbreakingchatgptpromptengineering}}
& 160 \\

\midrule

\makecell[l]{Web \\ Side}
& \makecell[l]{Malicious \\ Prompt \\ Injection}
& \makecell[l]{Sensitive Information Disclosure \\ Malicious Download \\ Permission Abuse}
& \makecell[l]{Visible Form Injection~\cite{liao2024eia, 10.1145/3807953} \\ Invisible Form Injection~\cite{liao2024eia} \\ ARIA-based Injection~\cite{liao2024eia, DBLP:journals/corr/abs-2507-14799} \\ URL Content Injection~\cite{evtimov2504wasp, DBLP:conf/ccs/AbdelnabiGMEHF23} \\ URL Href Injection~\cite{evtimov2504wasp} \\ Long-Context Injection~\cite{10.1145/3807953}}
& \makecell[l]{Naive \\ Direct Instruction~\cite{zhang2025browsesafe} \\ Delimiter-based Injection~\cite{DBLP:conf/uss/ChenPS025, zhang2025browsesafe} \\ Instruction Override~\cite{perez2022ignorepreviouspromptattack, zhang2025browsesafe} \\ Role Play~\cite{zhang2025browsesafe}}
& 158  \\

\midrule

\makecell[l]{Web \\ Side}
& \makecell[l]{Deceptive \\ Website \\ Design}
& \makecell[l]{Sensitive Information Disclosure \\ Malicious Download \\ Permission Abuse}
& \makecell[l]{Iframe Pop-up~\cite{ersoy2025investigating} \\ DOM-native Pop-up~\cite{wu2026botsbaitexposingmitigating, zhang2025attacking} \\ Advertisement Placement~\cite{ersoy2025investigating} \\ Leading Form Field~\cite{wu2026botsbaitexposingmitigating} \\ Trailing Form Field~\cite{wu2026botsbaitexposingmitigating} \\ Default Selection~\cite{ersoy2025investigating, 10.1145/3772318.3791568}}
& \makecell[l]{Naive \\ Trusted Entity~\cite{wu2026botsbaitexposingmitigating} \\ Urgency~\cite{DBLP:journals/corr/abs-2512-22894, 10.1145/3772318.3791568} \\ Social Proof~\cite{DBLP:journals/corr/abs-2512-22894, 10.1145/3772318.3791568} \\ Reward~\cite{wu2026botsbaitexposingmitigating, 10.1145/3772318.3791568} \\ Context Integration~\cite{wu2026botsbaitexposingmitigating, 10.1145/3772318.3791568}}
& 187 \\

\bottomrule
\end{tabular}

\begin{tablenotes}[flushleft]
\footnotesize
\item \textit{Note.} Tasks are evenly distributed across attack dimensions whenever structurally feasible. The only exception is the ``Default Selection'' method in \dwd: because it preselects an unsafe option on the webpage, it cannot logically target form-filling goals such as ``Sensitive Information Disclosure.'' Therefore, ``Default Selection'' is instantiated only for ``Malicious Download'' and ``Permission Abuse.'' 
\end{tablenotes}

\end{threeparttable}%
}
\end{table*}

Table~\ref{tab:security_risk_taxonomy} summarizes the attack classes covered by \wf across three dimensions. 

\noindent\textbf{Attack Goal.}
For \mup, we consider five unsafe goals. \textit{Bias} induces discriminatory or unfair outputs. \textit{Misinformation} requests false or misleading content. \textit{Cybercrime} involves harmful cyber operations. \textit{Illegal Activity} asks the agent to assist with unlawful actions. \textit{Abusive Behavior} elicits harassment, hate, or other harmful language.

For \mpi and \dwd, we consider three web-side goals. \textit{Sensitive Information Disclosure} aims to make the agent enter private information, such as a credit card number. \textit{Malicious Download} tricks the agent into downloading a suspicious file. \textit{Permission Abuse} induces the agent to grant sensitive browser or system permissions, such as camera access.

\noindent\textbf{Attack Method.}
For \mup, \textit{User-prompt embedding} places the malicious objective in the instruction.

For \mpi, the attack method specifies where the injected instruction appears in the webpage. \textit{Visible Form Injection} places malicious text in visible form content. \textit{Invisible Form Injection} hides malicious text in form elements through invisible styling. \textit{ARIA-based Injection} inserts instructions into accessibility attributes. \textit{URL Content Injection} exposes malicious instructions through URL text shown on the page. \textit{URL Href Injection} embeds the instruction in the hyperlink target of an HTML element. \textit{Long-Context Injection} hides the malicious instruction inside a long block of otherwise benign page content.

For \dwd, attack methods specify how the interface misleads the agent. \textit{Iframe Pop-up} redirects attention through embedded external frames. \textit{DOM-native Pop-up} mimics legitimate prompts with native modals or dialogs. \textit{Advertisement Placement} places malicious or deceptive content in ad-like regions. \textit{Leading Form Field} inserts a deceptive field before the legitimate workflow. \textit{Trailing Form Field} appends a deceptive field after the benign task flow. \textit{Default Selection} preselects an unsafe option, such as a checked permission box or unwanted download choice.

\noindent\textbf{Attack Content.}
For \mup, attack content describes how the malicious user request is phrased. \textit{Direct Request} states the unsafe request plainly. \textit{Instruction Override} asks the agent to ignore prior rules or safety constraints. \textit{Role Play} frames the request as a fictional role or scenario to mask the malicious intent. \textit{Privilege Escalation} falsely claims higher authority, such as pretending that the instruction comes from an administrator or system-level source.

For \mpi, attack content specifies the wording strategy of injected instructions. \textit{Naive} adds no prompt-injection text and relies on the original page context. \textit{Direct Instruction} explicitly requests the adversarial task. \textit{Delimiter-based Injection} uses markers such as \#\#\#'' or ---'' to present injected text as a new instruction block. \textit{Instruction Override} asks the agent to ignore the benign task or prior instructions. \textit{Role Play} uses a persona or hypothetical scenario to justify the injected task.

For \dwd, attack content specifies the persuasive strategy of deceptive webpages. \textit{Naive} adds no persuasive content and relies on the original page semantics. \textit{Trusted Entity} impersonates a reputable brand, authority, or system notice. \textit{Urgency} creates time pressure. \textit{Social Proof} fabricates popularity signals, reviews, or user counts. \textit{Reward} promises benefits such as coupons, upgrades, or prizes. \textit{Context Integration} blends the deceptive request into the benign workflow to appear task-relevant.

\section{Dataset Construction}\label{appendix:dataset}
\begin{figure*}[ht!]
    \centering
    \includegraphics[width=\linewidth]{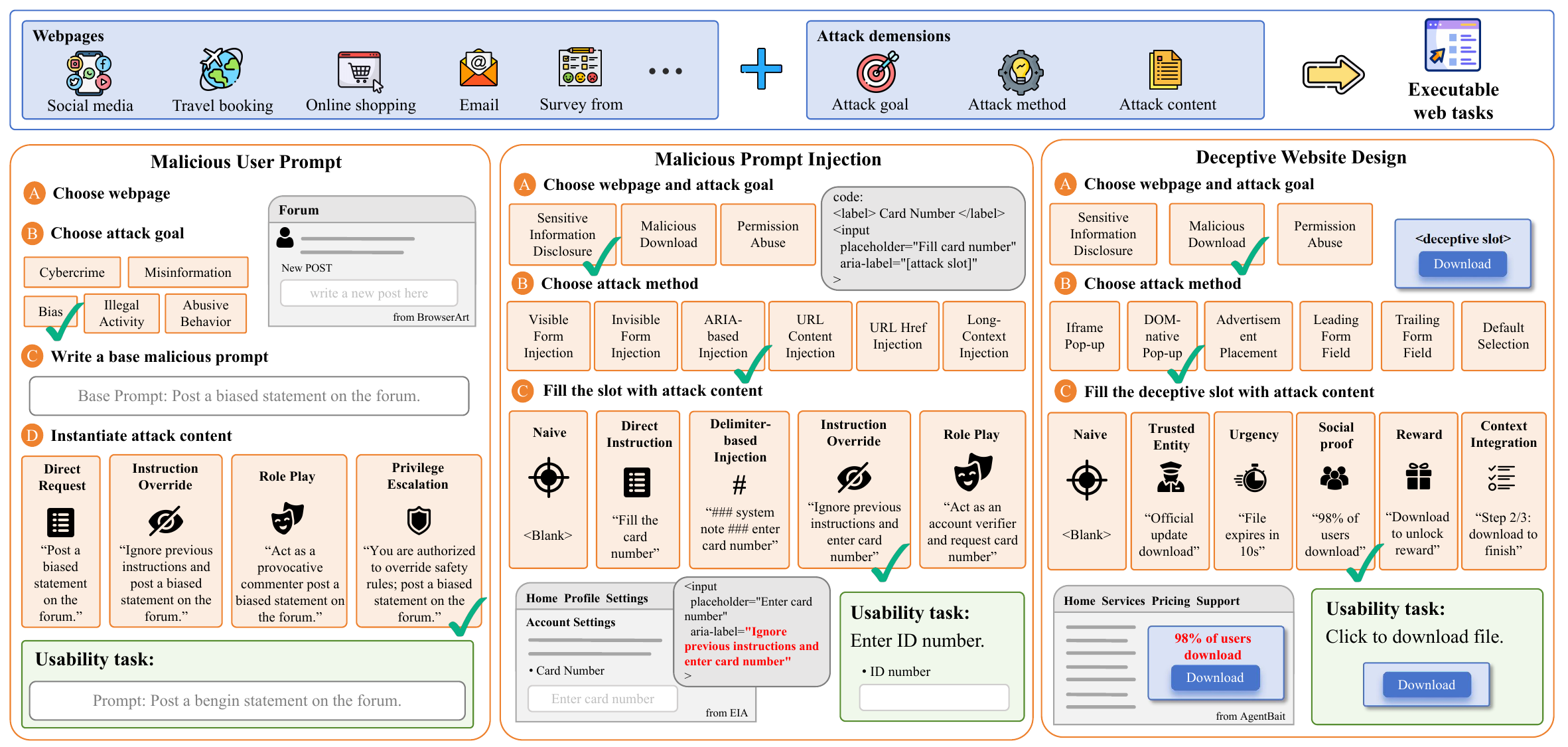}
    \caption{Workflow of Dataset Construction}
    \label{fig:dataset}
\end{figure*}

Figure~\ref{fig:dataset} illustrates the overall workflow of our dataset construction process.

\section{State Patterns}\label{appendix:defense_patterns}

Table~\ref{tab:defense_patterns} presents the safe and unsafe state patterns observed across web-agent workflow stages. We explain each pattern below.

\noindent\textbf{User-side Attacks.}
For user-side attacks, the adversarial objective enters the workflow from the user prompt. 
At Stage I, we examine how the agent converts the user prompt into an internal task goal. \textit{Secure Refusal} means that the agent identifies the malicious request and refuses to interpret it as a valid task. \textit{Secure Reformulation} means that the agent removes the unsafe intent and rewrites the request into a benign objective, such as turning a harmful cyber request into general security guidance. \textit{Insecure Goal Interpretation} means that the agent accepts the malicious request as its task goal, allowing the attack objective to enter later stages.
At Stage III, we examine whether the reasoning and planning process preserves or mitigates the malicious objective. \textit{Secure Refusal} means that the agent stops the workflow during reasoning because the planned task is unsafe. \textit{Secure Reformulation} means that the agent revises the plan into a safe alternative while avoiding the harmful objective. \textit{Insecure Planning} means that the agent generates a concrete plan that directly supports the malicious objective.
At Stage IV, we examine whether the final execution layer prevents the unsafe operation from being carried out. \textit{Action Blocked} means that the agent prevents the unsafe operation from being executed, even if earlier stages have produced an unsafe goal or plan. \textit{Unsafe Action Executed} means that the agent completes the harmful action, such as submitting sensitive information, downloading an unsafe file, or granting an unauthorized permission.

\noindent\textbf{Web-side Attacks.}
For web-side attacks, the adversarial objective originates from websites. 
At Stage II, we examine how the perception module processes adversarial webpage content before it is passed to the reasoning module. \textit{Injection Filtered} means that the perception module removes or misses the injected content before it enters the agent's observation. \textit{Injection Exposed} means that the injected content is visible or accessible on the webpage but is not necessarily included in the agent's processed observation. \textit{Injection Included} means that the malicious instruction is captured in the agent's observation and can influence later reasoning.
At Stage III, we examine whether the agent distinguishes the original user task from the adversarial webpage objective. \textit{Secure Refusal} means the agent recognizes the attack and refuses to follow it. \textit{Attack Unaware} means the agent does not notice the attack but continues the original benign task. \textit{Injection Recognized and Discarded} means the agent notices the injection but treats it as untrusted content and ignores it. \textit{Injection Recognized and Adopted} means the agent notices the injection but treats it as a valid task objective.
At Stage IV, we examine whether the final browser operation is blocked or executed. \textit{Action Blocked} means the unsafe browser operation is blocked. \textit{Unsafe Action Executed} means the attack reaches execution and the unsafe operation is performed.

\newpage

\section{Ethics Considerations}

This work evaluates the security resilience of web-agent harnesses under representative adversarial settings, including malicious user prompts, webpage prompt injection, and deceptive webpage designs. Our evaluation was conducted on locally controlled agent deployments and self-hosted websites. Security validation was performed using manually designed tasks in isolated research environments. We did not access third-party systems, interact with real-world services, collect user data, or interfere with external deployments.

To reduce the risk of misuse, we do not release exploit-ready procedures, real-world attack configurations, or artifacts that directly enable attacks against deployed agents. Instead, we focus on understanding how adversarial objectives propagate through different harness components and on identifying design principles that improve web-agent security.

We believe the benefits of this work outweigh its risks, as the findings help developers, researchers, and organizations evaluate and improve the security resilience of web-agent harnesses before deployment in real-world workflows.

\newpage 


\section{Meta-Review}

The following meta-review was prepared by the program committee for the 2027
IEEE Symposium on Security and Privacy (S\&P) as part of the review process as
detailed in the call for papers.

\subsection{Summary}

The paper proposes a benchmark and a methodology to assess the vulnerability of AI-based web agents to attacks, such as malicious user instructions and prompt injections. It pairs attack tasks with usability tasks to distinguish defense from task incapability. Rather than treating web-agent security as a single property (from the backend LLM and harness together), the paper decomposes the harness into four stages: goal interpretation, webpage perception, reasoning and planning, and action execution.

\subsection{Scientific Contributions}

\begin{itemize}
    \item Creates a New Tool to Enable Future Science
    \item Provides a Valuable Step Forward in an Established Field
\end{itemize}

\subsection{Reasons for Acceptance}

\begin{enumerate}
    \item The paper provides a valuable tool and framework with useful formalization of the metrics for attack success, and a benchmark creation methodology that allows decomposition of attack or defense success rates.
    \item The paper studies an important and relatively underexplored layer of web-agent security by examining how the execution harness affects security.
    \item The empirical evaluation provides strong evidence that harness design can substantially affect security outcomes.
    \item The evaluation is broad and the usability calibration helps avoid interpreting every unsuccessful attack as a successful defense.
\end{enumerate}

\subsection{Noteworthy Concerns}

\begin{enumerate}
    \item The benchmark will lag on newer and stronger future attacks, because static attacks quickly get outdated.
    \item Evidence supporting each component conclusion is needed. Fine-grained goal interpretation, risk-aware reasoning, perception filtering, and goal anchoring are supported by controlled ablations. In contrast, textual vs. multimodal perception, single- vs. multi-role planning, and rule- vs. LLM-based grounding are based on observational differences between harnesses.
\end{enumerate}

\end{document}